\documentclass[prd,superscriptaddress,altaffilletter,showpacs,nofootinbib]{revtex4}
\usepackage{amssymb}
\usepackage{amsmath}
\usepackage{graphicx}
\usepackage{subfigure}
\usepackage{epsfig}
\usepackage{epstopdf}
\usepackage[usenames,dvipsnames]{color}
\usepackage{dcolumn}
\usepackage{bm}

\begin{document}
\title{Localization of bulk matter fields, the hierarchy problem and 
corrections to Coulomb's law on a pure de Sitter thick braneworld}

\author{Heng Guo}\email{guoheng81@gmail.com}
\affiliation{School of Science, Xidian University, Xi'an 710071, P. R. China}
\affiliation{Institute of Theoretical Physics, Lanzhou University, Lanzhou 730000, P. R. China}
\author{Alfredo Herrera--Aguilar}\email{aha@fis.unam.mx}
\affiliation{Instituto de Ciencias F\'{\i}sicas, Universidad Nacional Aut\'onoma de M\'exico,\\
Apartado Postal 48-3, MEX--62251, Cuernavaca, Morelos, M\'{e}xico}
\affiliation{Instituto de F\'{\i}sica y Matem\'{a}ticas, Universidad Michoacana de San Nicol\'as
de Hidalgo,\\Edificio C--3, Ciudad Universitaria, CP 58040, Morelia, Michoac\'{a}n, M\'{e}xico}
\author{Yu--Xiao Liu}\email{liuyx@lzu.edu.cn}
\affiliation{Institute of Theoretical Physics, Lanzhou University, Lanzhou 730000, P. R. China}
\author{Dagoberto Malag\'on--Morej\'on}\email{malagon@fis.unam.mx}
\affiliation{Instituto de Ciencias F\'{\i}sicas, Universidad Nacional Aut\'onoma de M\'exico,\\
Apartado Postal 48-3, MEX--62251, Cuernavaca, Morelos, M\'{e}xico}
\affiliation{Instituto de F\'{\i}sica y Matem\'{a}ticas, Universidad Michoacana de San Nicol\'as
de Hidalgo,\\Edificio C--3, Ciudad Universitaria, CP 58040, Morelia, Michoac\'{a}n, M\'{e}xico}
\author{Refugio Rigel Mora--Luna},\email{rigel@ifm.umich.mx}
\affiliation{Instituto de F\'{\i}sica y Matem\'{a}ticas, Universidad Michoacana de San Nicol\'as
de Hidalgo,\\Edificio C--3, Ciudad Universitaria, CP 58040, Morelia, Michoac\'{a}n, M\'{e}xico}

\date{\today}

\begin{abstract}
In this paper we investigate the localization and mass spectra of various matter fields with spin
$0$, $1$ and $1/2$ on a geometric thick brane generated by pure 4D and 5D positive cosmological
constants without bulk scalar fields. This model possesses a 4D cosmological constant that can be
made as small as one desires without fine-tuning it with the bulk cosmological constant. The
Randall--Sundrum model was obtained as an analytic continuation of the flat brane limit of this
braneworld configuration when the Hubble parameter disappears. We also show that within the
framework of this inflating braneworld model it is possible to formulate a mechanism for obtaining
TeV mass scales from Planck ones by adding a {\it positive} thin brane, where the Standard Model
fields are trapped, located at a distance $y_2$ from the origin, where the Planck thick brane
resides. The brane separation must be of the {\it same order} than the inverse thickness parameter
of the model in order for the mechanism to generate the desired hierarchy. This result is obtained
by imposing the recovery of both the correct 4D gravitational couplings and the actually observed
accelerated expansion of the universe in our de Sitter braneworld. Regarding the localization of
matter in the purely geometric thick braneworld, for spin $0$ massless and massive scalar fields
as well as for spin $1$ vector fields, the potentials of the Kaluza--Klein (KK) modes in the
corresponding Schr\"{o}dinger equations are modified P\"{o}schl--Teller potentials, which lead to
the localization of the scalar and vector zero modes on the brane as well as to mass gaps in the
mass spectra. We also compute the corrections to Coulomb's law coming from massive KK vector modes. 
For spin $1/2$ fermions, we introduce the bulk mass term $MF(z)\bar{\Psi}\Psi$ in
the action and four different cases are investigated. Localization of the massless left--chiral
fermion zero mode is feasible for just two cases of $F(z)$. In the first one we obtain
Schr\"{o}dinger equations with modified P\"{o}schl--Teller potentials with the corresponding mass
gaps for both left-- and right--chiral fermions, the number of massive KK bound states is {\it
finite} (determined by the ratio $M/b$) and is the same for left-- and right--chiral modes. In the
second case we get a family of solutions with an {\it infinite} number of bound states where the
mass spectra for left-- and right--chiral KK fermion modes are discrete. A special case resembles
the one--dimensional quantum harmonic oscillator problem. For both of these cases, the mass
spectrum of right--chiral fermions is shifted with respect to the mass spectrum of the
left--chiral ones.
\end{abstract}

\pacs{11.25.Mj, 04.40.Nr, 11.10.Kk} \keywords{Large Extra Dimensions, Field Theories in Higher
Dimensions, Hierarchy problem}



\maketitle

\section{Introduction}
The idea that our observed four--dimensional (4D) universe can be considered as a brane, which is
embedded in a higher dimensional space--time, has received considerable attention. In the
string/M--theory context, branes naturally appear and provide a novel mode for discussing
phenomenological and cosmological issues related to extra dimensions. The proposal that extra
dimensions may be non--compact
\cite{RubakovPLB1983136,VisserPLB1985,Randjbar-DaemiPLB1986,gog,rs,Lykken} or large
\cite{AntoniadisPLB1990,ADD} can supply new insights for solving the gauge hierarchy problem
\cite{gog,ADD} and the cosmological constant problem
\cite{RubakovPLB1983136,Randjbar-DaemiPLB1986,CosmConst}. The framework of brane scenarios is that
gravity is free to propagate in all dimensions, however all 4D matter fields (electromagnetic,
Yang--Mills, etc.) are confined to a 3--brane with no contradiction with present gravitational
experiments \cite{RubakovPLB1983136,VisserPLB1985,ADD}. In Refs. \cite{gog,rs}, an alternative
scenario, the so-called Randall--Sundrum (RS) braneworld model showed that the effective
4D gravity could be recovered even in the case of non--compact extra dimensions.

A more natural theory arises in the framework of thick brane scenarios, which are generally based
on gravity coupled to one or several scalars
\cite{De_Wolfe_PRD_2000,PRD_Stojkovic,Gremm_2000,Giovanninietal,Csaki_NPB_2000,CamposPRL2002,
dSbw,Sfetsos,varios,ThickBrane,BGL,Liu0907.1952,Guo_jhep,BajcPLB2000,
OdaPLB2000113,Liu0708,NonLocalizedFermion,NonLocalizedFermion2,Volkas0705.1584,
IchinosePRD2002,Ringeval,GherghettaPRL2000,Neupane,RandjbarPLB2000,
KoleyCQG2005,DubovskyPRD2000,0803.1458,LiuJHEP2007,0901.3543,
Liu0907.0910,Koley2009,0812.2638,ThickBraneWeyl,Cvetic,PRD0709.3552,ThickBrane4,LiuJCAP2009,Liu0803,
KobayashiPRD2002,RandallJHEP2001,0910.0363,
afonso_plb2006,liu_0911.0269,liu_0909.2312,1004.0150,Liu_2010,0610052,1009.1684,1012.4534}. In
these scenarios, the scalar fields do not play the role of bulk fields, but provide the
``material" from which the thick branes are made of. For some comprehensive reviews about thick
branes please see Refs. \cite{0812.1092,0904.1775,0907.3074,1003.1698,brane_book,1004.3962}. The
majority of these scalar thick brane models has a continuous KK graviton spectrum starting at zero
mass, causing a need for explaining why such arbitrarily light extra gravitons have not led to
detectable deviations from standard gravity. Although there are some mechanisms that try to
explain how such light extra dimensional gravitons could go unobserved (see \cite{Lykken}, for
instance), it is interesting to enquire whether thick brane solutions with a mass gap in the
graviton spectrum exist and thereby avoid this problem as well. It turns out that the existence of
a mass gap in the spectrum of KK fluctuation modes provides an easy way to control the excitation
of the KK gravitons and the concomitant energy loss into the fifth dimension. However, for thick
braneworld models with 4D Poincar\'e symmetry, the existence of a mass gap involves the presence
of naked singularities at the boundaries of the relevant manifold \cite{0910.0363}, a problem that
can sometimes be healed with the introduction of a 3--brane with de Sitter ($dS$) symmetry
\cite{dSbw} or non standard branes \cite{Sfetsos}.

In braneworld scenarios, there is an interesting issue: whether various bulk fields could be
confined to the brane by a natural mechanism. Generally, massless scalar fields \cite{BajcPLB2000}
and gravitons \cite{rs} can be localized on branes of different types. Vector fields can be
localized on the RS brane in some higher--dimensional cases \cite{OdaPLB2000113} or on thick $dS$
branes and Weyl thick branes \cite{Liu0708}. The feature of fermion localization is very important
for these models. In general, if one does not introduce the coupling between the fermions and bulk
scalars, the fermions do not have normalizable zero modes in five and six dimensions
\cite{BajcPLB2000,OdaPLB2000113,Liu0708,NonLocalizedFermion,IchinosePRD2002,Ringeval,
GherghettaPRL2000,Neupane,RandjbarPLB2000,KoleyCQG2005,DubovskyPRD2000,0803.1458,
LiuJHEP2007,0901.3543,Liu0907.0910,Koley2009,0812.2638}. In some cases, with scalar--fermion
coupling, a single bound state and a continuous gapless spectrum of massive fermion KK states can
be obtained \cite{Liu0708}. In some other brane models, one can obtain finite discrete KK states
(with a mass gap between them) and a continuous gapless spectrum starting at a positive $m^2$
\cite{LiuJCAP2009,Liu0803}.

Recently, a pure de Sitter thick brane with very appealing features has been investigated in Ref.
\cite{1009.1684}. This thick brane is not made of bulk scalars, but rather is modeled by an
intriguing relation between the curvatures generated by the 4D and 5D
cosmological constants without the inclusion of bulk scalar fields at all. Thus, this regular
thick brane configuration emerges as a nonlinear solution to the full Einstein equations of a
model which describes five--dimensional (5D) gravity with a positive cosmological constant where the 3--brane
possesses de Sitter symmetry instead of the Poincar\'e one. Furthermore, a finite effective 4D
Planck mass (which also depends on the scale factor parameter of the 3--brane: $M_{Pl}^2=\frac{\pi
M_*^3 H^2}{2b^3}$, where $M_*^3=(2\kappa_5)^{-2}$) was obtained in this model upon integration of
the fifth dimension; the localization of 4D gravity with the correct couplings takes place in such a way 
that there exists a mass gap between the 4D stable massless graviton and the massive KK excitation 
modes, a useful peculiarity of the model from the phenomenological viewpoint. Finally, the 
corresponding corrections to Newton's law arising in this model have already been computed in analytic 
form \cite{1009.1684}.

One of the advantages of this regular $dS_4$ braneworld model in comparison to thick branes
generated by scalar fields is that the latter develop naked singularities at the boundaries of the
fifth dimension when the presence of a mass gap in the spectrum of the 5D KK graviton fluctuations
is required \cite{De_Wolfe_PRD_2000,Gremm_2000,ThickBrane4,0910.0363}. Moreover, the bulk scalar
fields themselves usually are divergent whithin the framework of these braneworlds. The existence
of a mass gap is quite relevant from the phenomenological point of view since in this case the
massless physical 4D graviton is separated from the massive KK modes, a fact that fixes the energy
scale at which these massive fluctuations can be excited and enables us to avoid difficulties when
analyzing the traces of ultra light KK excitations
\cite{De_Wolfe_PRD_2000,Gremm_2000,PRD0709.3552}. This kind of mass gaps is present in braneworld 
models with a de Sitter metric induced on the brane like in \cite{dSbw}, where, however, the authors have
made use of delta function sources which lead to singularities at the position of the branes. 

Thus, within this regular braneworld one is able to model the thick brane geometrically avoiding
the use of scalar matter at all, in which the 4D gravity can be localized, and as an extra bonus,
a mass gap is displayed in the gravity spectrum of KK excitations without developing naked
singularities as in the case of scalar thick brane configurations.

Another interesting feature of this thick braneworld model is that its 4D cosmological constant
can be made as small as one desires without the need of a fine-tuning with the bulk cosmological
constant, as it happens in the Randall--Sundrum model (see Sec. \ref{SecModel} for more details).

A last but not less important advantage of this model is that it allows us to analytically study
the behavior of the massive modes of the spectrum of KK excitations, a scarce phenomenon when
considering smooth brane configurations, since usually the corresponding Schr\"{o}dinger equation
cannot be explicitly integrated for $m$ greater than zero and one is forced to make use of
numerical analysis (see \cite{De_Wolfe_PRD_2000,Gremm_2000,BGL}, for instance).

This kind of embedded solutions are also interesting from the viewpoint of cosmology. Since this
solution describes a 3--brane with de Sitter symmetry (expanding, maximally symmetric space, or,
$dS_4$) embedded in a $dS_5$ space, it mimics some aspects of the inflationary period of our
Universe \cite{guth,linde}. It turns out that the Cosmic Inflation theory is in good agreement
with the properties of the temperature fluctuations seen in the Cosmic Microwave Background
Radiation. This and other observational successes, together with the fact that inflation likely
takes place at very high temperatures and its study involves making assumptions about the relevant
physics that takes place at such high energies, lead to several attempts to find inflationary
configurations within string theory and supergravity \cite{popeetal,burgess,SC}. Moreover, since
in this braneworld solution the second derivative of the scale factor with respect to time is
positive, it also can describe some aspects of the accelerated expansion of the Universe, related
to its present epoch.

One more interesting issue in which this kind of configurations can have physical applications is
the realization of the dS/CFT correspondence project \cite{strominger1,strominger2}, where having
exact de Sitter vacua will be of outstanding relevance for understanding such a correspondence
and, hence, to get insight into the quantum gravity nature of de Sitter space. Despite the
difficulties that this project faces (like the observer dependence nature of the de Sitter event
horizon \cite{LesHouchesdS} and the lack of supersymmetry for de Sitter space \cite{PNS}, in
contrast with its AdS/CFT counterpart), a concrete example has been worked out for lower
dimensional cases \cite{guijosaetal}.

In this paper we will explore the localization and mass spectra of various matter fields with spin
$0$, $1$ and $1/2$ on this thick brane scenario. For the spin $0$ scalars and the spin $1$
vectors, the zero modes can be localized on the brane, and there exists a mass gap in their mass
spectra. However, the localization of the spin $1/2$ fermions is very special since there is no
scalar field to couple with in this model, in contrast to thick branes generated by scalar fields.
This issue has been previously treated by several authors (see
\cite{JackiwPRD1976,BajcPLB2000,NonLocalizedFermion,OdaPLB2000113,RandjbarPLB2000,PRD_Oda_026002},
for instance), but in fact, the mechanism for localization of spin $1/2$ fermions on a brane (with
the aid of a mass term generated by the Higgs field with a vacuum expectation value of a kink
profile) was proposed for the first time in \cite{JackiwPRD1976} for a flat spacetime,
subsequently extended to the $AdS_5$ case in \cite{BajcPLB2000,NonLocalizedFermion} and further
generalized to higher--dimensional spacetimes in \cite{RandjbarPLB2000}. Thus, in order to trap
the fermions, in this work we introduce the bulk mass term $MF(z)\bar{\Psi}\Psi$, which contains
a sort of spread mass along the fifth dimension, so that the character of the localization is
different for different forms of the mass function $MF(z)$.

The organization of this paper is as follows: In Sec. \ref{SecModel}, we first review the so--called pure 
de Sitter thick brane and present some new aspects of this model like its flat (static) limit $H\rightarrow 0$, 
the possibility of having an effective 4D cosmological constant as small as desired, as well as a mechanism 
that allows one to obtain TeV mass scales from Planck ones on a
Standard Model thin brane located at a certain distance from the Planck brane where gravity is
localized. Then, in Sec. \ref{SecLocalization}, we investigate the localization and mass spectra
of various matter fields with spin $0$, $1$ and $1/2$ on this thick brane. We further compute the
corrections to Coulomb's law due to the presence of massive KK vector modes in Sec. \ref{CCL}. 
Finally, our conclusion is given in Sec. \ref{SecConclusion} together with some discussion on the 
presented material.

\section{Review of the thick braneworld generated by pure curvature} 
\label{SecModel}

As mentioned above, an intrinsic appealing feature of our regular de Sitter braneworld is that it
does not need to include a scalar field nor any other kind of bulk matter in order to smooth out
the delta function singularities of the Randall-Sundrum model. Thus, this thick braneworld arises
in a completely geometric way: it is just the curvature that localizes gravity and other matter
fields on the regular 3--brane. Thus, we start with the following 5D action for a
thick braneworld model \cite{1009.1684}
\begin{equation}
 S= \frac{1}{2\kappa_5^2}\int d^5 x \sqrt{-g}( R - 2\Lambda_{5}),
\label{action}
\end{equation}
where $R$ is the 5D scalar curvature, $\Lambda_{5}$ is the bulk cosmological
constant, and $\kappa_5^2=8\pi G_5$ with $G_5$ the five--dimensional Newton constant. Here we use
the signature $(-++++)$ and the Riemann tensor, defined as follows
$R_{MNPQ}=\frac{\Lambda_5}{6}\left(g_{MP}g_{NQ}-g_{MQ}g_{NP}\right)$, gives rise to the Ricci one
$R_{NQ}=R^M_{NMQ}$ upon contraction of its first and third indices, where $M,N,P,Q=0,1,2,3,5.$
From this action, the Einstein equations with a cosmological constant in five dimensions are easy
to get
\begin{eqnarray}\label{EinsteinEq_5d}
R_{MN}-\frac{1}{2}R\ g_{MN}=-\Lambda_{5}\ g_{MN}.
\end{eqnarray}
The most general 5D metric compatible with an induced 3--brane with spatially flat
cosmological background can be taken to be
\begin{eqnarray}\label{metric_y}
 ds^2 &=& g_{MN}dx^{M}dx^{N}
         =\text{e}^{2A(y)}\left[\hat{g}_{\mu\nu}(x)dx^\mu dx^\nu \right]+ dy^{2}
         \nonumber \\
      &=& \text{e}^{2A(y)}\left[ -dt^{2}+ a^{2}(t)
             (dx_{1}^{2}+dx_{2}^{2}+dx_{3}^{2})\right]
             +dy^{2},
\end{eqnarray}
where $\text{e}^{2A(y)}$ and $a(t)$ are the warp factor and the scale factor of the brane,
$\hat{g}_{\mu\nu}(x)$ denotes the 4D metric tensor ($\mu,\nu=0,1,2,3$) and $y$ stands for the
extra dimensional coordinate.

By considering the metric (\ref{metric_y}) we compute the Einstein equations which reduce to a 
very simple system \cite{1009.1684}\footnote{In the original work there is a misprint 
in Eq. (\ref{EinsteinEq_b}): the factor $\text{e}^{-2A(y)}$ is multiplying the whole expression in 
brackets, however, it should multiply just the expression within the parenthesis, as quoted here.}:
\begin{subequations}\label{EinsteinEq}
\begin{eqnarray}
 \label{EinsteinEq_a}
 A''(y)&=& \frac{1}{3}\left(\frac{2\dot{a}^{2}}{a^{2}}-\frac{5\ddot{a}}{a} \right)
           \text{e}^{-2A(y)}, \\
 \label{EinsteinEq_b}
 A'^{2}(y)&=& \frac{1}{6}\left[ \left( \frac{5\ddot{a}}{a}+\frac{\dot{a}^{2}}{a^{2}} \right)
               \text{e}^{-2A(y)}-\Lambda_{5}\right],
\end{eqnarray}
\end{subequations}
where the prime and the dot denote derivative with respect to $y$ and $t,$ respectively. 
It turns out that the equations for $A$ and $a$ decouple in this system and
the second order differential equation for the scale factor reads
\begin{eqnarray}\label{EqScaleFactor}
a\ddot{a}-\dot{a}^{2}=0.
\end{eqnarray}
The general solution of Eq. (\ref{EqScaleFactor}) is
$a(t)=a_0\text{e}^{Ht}$ with $a_0$ and $H$ being arbitrary constants, and
we can choose the scale factor corresponding to a de Sitter
4D cosmological background
\begin{eqnarray}\label{solution_a}
a(t)=\text{e}^{Ht},
\end{eqnarray}
since the constant $a_0$ can be absorbed into a coordinate redefinition. Here $H$ is the Hubble
parameter and $3H^{2}=\Lambda_{4}$ with $\Lambda_{4}$ being the effective 4D
cosmological constant obtained upon integration of the fifth dimension \cite{brane_book}, a result
which also is consistent with the computation performed from the viewpoint of the induced 4D
gravity on the brane by making use of the Shiromizu--Maeda--Sasaki formalism which projects the 5D
curvature along the 3--brane \cite{SMS,1004.3962}. On the other hand, the solution for the warp 
factor is found to be \cite{1009.1684}:
\begin{eqnarray}\label{warpfactor} A(y)=\ln\left[\frac{H}{b}\cos(b(y-y_0))
\right],
\end{eqnarray}
where $y_0$ is a constant that labels the position of the brane, and $1/b$ parameterizes the thickness 
of the 3--brane and is related to the 5D cosmological constant as follows:
\begin{eqnarray}\label{L5}
\Lambda_{5}=6b^{2}.
\end{eqnarray}
Thus, the relevant 5D metric with an induced de Sitter 3--brane reads
\begin{eqnarray}\label{metric_yfull}
 ds^2 = \frac{H^2}{b^2}\cos^2(by)\left[ -dt^{2}+ \text{e}^{2Ht}
             (dx_{1}^{2}+dx_{2}^{2}+dx_{3}^{2})\right]
             +dy^{2}.
\end{eqnarray}
From this form of the warp factor, we come to the following conclusion: a thick de Sitter brane is
localized around $y_0=0$, and the range of the fifth dimension is $-\left|\frac{\pi}{2b}
\right|<y<+ \left|\frac{\pi}{2b}\right|$. The localization and stability properties of graviton
fluctuations on this brane were studied in \cite{1009.1684}.

This metric arises as a solution where both cosmological constants $\Lambda_4$ and $\Lambda_5$ are
positive and, hence, accounts for the embedding of a $dS_4$ 3--brane into a chart of the $dS_5$
space--time, a result in agreement with \cite{neupane3}, where it was pointed out that an $AdS_5$
bulk in itself may be problematic in the absence of other bulk fields.

Our purely geometric braneworld solution requires both $\Lambda_4$ and $\Lambda_5$ to be nontrivial. 
Setting one of them alone to zero leads to unphysical solutions. However, since the solution possesses 
two arbitrary parameters, we can perform a double limit in order to obtain a physically meaningful 
solution as we shall see below.

\subsection{A flat thin brane limit}

An interesting issue is to see whether there is a meaningful limit when the Hubble parameter
vanishes $H=0$. In order to clarify this point, we can also express the solution of the Einstein
equations (\ref{EinsteinEq}) in the following way:
\begin{eqnarray}
A(y)=\ln
 \left[\text{e}^{-iby}\frac{H^{2}+\text{e}^{2iby}\left(b-\sqrt{b^2-H^2}\right)^{2}}
        {2b\left(b-\sqrt{b^2-H^2}\right)}
 \right],
\end{eqnarray}
which is equivalent to (\ref{warpfactor}) with $by_0=i\ln\left(\frac{b-\sqrt{b^2-H^2}}{H}\right)$.
When the parameter $H$ is small, the above solution can be expressed as
\begin{eqnarray}
A(y)\backsimeq \ln
 \left[\text{e}^{-iby}\frac{H^{2}+\text{e}^{2iby}\left(\frac{H^{2}}{2b}\right)^{2}}
        {2b(\frac{H^{2}}{2b})}
 \right]
 = \ln
 \left[\text{e}^{-iby}\bigg(1+\text{e}^{2iby}\left(\frac{H}{2b}\right)^{2}\bigg)
 \right].
\end{eqnarray}
Thus, by taking the limit
$H\rightarrow 0$, this solution will be reduced to the case of the flat 3--brane:
\begin{eqnarray}
A(y)=-iby\equiv -k |y|,
\end{eqnarray}
where we have performed an analytical continuation of the $b$ parameter: $ib\equiv k$ and imposed
$Z_{2}$--symmetry along the extra coordinate. Thus, as we see, this is not a standard limit and 
requires passing from a $dS$ bulk to a $AdS$ one, a sort of phase transition that must be studied further, 
perhaps within the context of an extended version of this minimalistic model and a more complete solution where the transition from a positive to a negative curvature bulk could be clarified. 

%
%
Thus, the analytic continuation of this limit yields the famous Randall--Sundrum solution
\cite{gog,rs}, which implies that we need to add to the 5D action the corresponding delta (thin)
branes in the 4D space--time for mathematical consistency. This result can be seen as well at the
level of the differential equations (\ref{EinsteinEq}) since if we set there $H=0$, we see from
(\ref{EinsteinEq_b}) that the constant $\Lambda_5$ must necessary be negative since it is
proportional to $-A'^{2}$; we further impose the $Z_{2}$--symmetry to get the solution
\begin{eqnarray}
 A(y)=\pm \sqrt{-\frac{\Lambda_{5}}{6}}~|y|,
\end{eqnarray}
which just corresponds to the RS brane solution. Hence, the conclusion is that a pure flat brane
must be embedded in 5D anti-de Sitter space time, and the flat brane must be a thin
brane, as in the RS braneworld, which has been extensively studied in the literature and thus, in
this paper we will not investigate further this model.

Since our universe is expanding in an accelerated way, it can be described by our $dS_4$ 3--brane
since $\ddot{a}>0$, a fact that is remarkable for the cosmology of our model. Moreover, as stated
above, the 4D effective cosmological constant is related to the Hubble parameter as usual, i.e.,
$\Lambda_{4}=3H^{2}$.

At this stage it is important to write the metric (\ref{metric_yfull}) in a conformal form by
performing the coordinate transformation
\begin{equation}\label{transformation}
dz=\text{e}^{-A(y)}dy.
\end{equation}
Then, the following expression for $z$ can be obtained
\begin{eqnarray}\label{zy}
z(y)=\int\text{e}^{-A(y)}dy=\frac{2}{H}\text{arctanh}
\left[\tan\left(\frac{by}{2}\right)\right].
\end{eqnarray}
It is easy to see that $z\rightarrow \pm\infty$ as
 $y\rightarrow \pm\left|\frac{\pi}{2b}\right|$, so the range of $z$
is $-\infty < z< +\infty$. Due to this transformation, the warp
factor $A$ can be rewritten as a function of $z$:
\begin{eqnarray}\label{Az}
A(z)=\ln\left[\frac{H}{b}\text{sech}(Hz)\right], \label{warpfactorz}
\end{eqnarray}
and the metric adopts the form:
\begin{eqnarray}\label{metric_z}
 ds^{2}&=& \text{e}^{2A(z)}\left[\hat{g}_{\mu\nu}(x)dx^\mu dx^\nu + dz^2 \right]
                \nonumber \\
       &=& \frac{H^2}{b^2}\text{sech}^2(Hz)\left[-dt^{2}+
                \text{e}^{2Ht}\left(dx_{1}^{2}+dx_{2}^{2}+dx_{3}^{2}\right)+dz^{2}\right].
\end{eqnarray}

An interesting feature of the above presented model is that, since we have two parameters in the
solution, $b$ and $H$, the 4D cosmological constant is completely independent of the 5D one. It
seems that in certain coordinate systems (where $b=H$) these two cosmological constants could be
related to each other (see \cite{popeetal}, for instance). However, by choosing a suitable
coordinate system, $\Lambda_4$ can be made as small as one wishes in an explicit way, while
keeping invariant the 5D cosmological constant and other 5D geometrical properties of the model.
However, it should be pointed out that by making small the value of $H$, the masses of the excited
KK particles also will adopt very small values as it will be discussed below in Sec. 
\ref{SecConclusion}.

\subsection{TeV vs Planck mass scales}

Another important issue within this de Sitter braneworld model consists of getting TeV physical
mass scales from fundamental masses of the order of the Planck scale. In order to study this
mechanism {\it \`a la} RS we should generalize our purely geometric 5D setup by adding in a
self--consistent way two thin branes and imposing $Z_2$--symmetry along the extra dimension. This
mathematically amounts to orbifolding the extra dimension, i.e. to replacing $y\longrightarrow
|y|$ and imposing periodicity along the fifth dimension when computing derivatives of $|y|$. This,
in turn, leads to junction conditions on each brane (labeled by $a=1,2$) to be fulfilled:
\begin{equation}
\lambda_a = (-1)^a\, b \tan\left(b\,y_a\right), \label{tensions}
\end{equation}
where $\lambda_a$ denotes the tension of the $a^{\rm th}$ brane and $y_a$ stands for their
positions. As we see, the first junction condition can be ignored by putting the Planck brane at
$y_1=0$, getting a tensionless configuration which is compatible with our purely geometric de
Sitter thick braneworld which localizes gravity. Moreover, the second junction condition leads to
a positive tension brane by setting $y_2>0$, avoiding at all the use of branes with negative
tension as in the RS model.

Thus, the thick Planck brane is located at the origin of the fifth dimension and localizes
gravity, while the thin TeV brane is located at a certain distance, $y_2$, from the first one and
hosts the Standard Model fields with electro--weak interactions. It can be shown that when
considering fundamental fields, like the Higgs field, in the TeV brane, the 4D physical mass
scales are determined by the following symmetry breaking scale
\begin{equation}
m = \frac{H}{b} \cos\left(b\,y_2\right) m_0, \label{m}
\end{equation}
after an appropriate field normalization.

Therefore, when the prefactor of $m_0$ is of order $10^{-15}$ we obtain TeV physical mass scales
from Planck ones. It is remarkable that this mechanism can indeed be established in the framework
of our model.

Let us recall that the effective 4D Planck mass of the thick de Sitter braneworld is finite along
the whole fifth dimension and reads \cite{1009.1684}:
\begin{equation}
M_{Pl}^2 = \frac{\pi}{2} \frac{M_*^3 H^2}{b^3}. \label{MPl1}
\end{equation}
By assigning the currently observed value to the Hubble parameter $H\approx 10^{-60}M_{Pl}$ and
setting $M_*\approx 10^{-15}M_{Pl}$ to consider TeV mass scales, we get $b\approx 10^{-55}M_{Pl}$
in order to recover the correct 4D gravitational couplings on the thick brane. Of course, this
assignment can be done in the understanding that we are fine tuning the value of the effective 4D
cosmological constant on the brane.

However, for the generalized setup we need the value of the effective 4D Planck mass between the
TeV and the Planck branes which is
\begin{equation}
M_{Pl}^2 = \frac{M_*^3 H^2}{b^3}\left[b\,y_2 + \frac{1}{2}\sin\left(2\,b\,y_2\right)\right].
\label{MPl2}
\end{equation}
This result shows that when $b\,y_2$ approaches the boundary of the orbifold, i.e. when
$b\,y_2\longrightarrow \pi/2$, the second term in (\ref{MPl2}) can be neglected in comparison with
the first one. This means that for large values of $b\,y_2$ the effective 4D Planck mass weakly
depends on the thin brane position $y_2$ and, hence, on the compactification radius of the extra
dimension of the manifold (see below).

It is also important to note that the $H\rightarrow 0$ limit of the formula (\ref{MPl1}) is well
defined since we can make $b$ tend to zero $b\rightarrow 0$ as well in such a way that the
following quotient approaches a finite constant $H/b^{3/2}\rightarrow C$, a fact that still leads
to a finite effective 4D Planck mass.

Therefore, the physically relevant ratio
\begin{equation}
\frac{m}{m_0} = 10^{-15}, \label{m2}
\end{equation}
can be achieved when
\begin{equation}
b\,y_2 = \arccos\left(\frac{m}{m_0}\frac{b}{H}\right) \approx \arccos\left(10^{-10}\right)
\lesssim \frac{\pi}{2}\approx 1, \label{by2}
\end{equation}
since $b/H\approx 10^{5}$ and $\arccos\left(10^{-10}\right)$ is slightly less than $\pi/2$.

Thus, is is remarkable that in order to generate the desired mass hierarchy we need a
compactification scale of the {\it same order} of the parameter $b$:
\begin{equation}
\mu_{c}\equiv \frac{1}{y_2}\approx b. \label{Mc}
\end{equation}
Moreover, while the required ratio to get TeV physical mass scales from fundamental Planck mass
scales is $m/m_0\approx 10^{-15}$, the necessary relevant ratio between the actually observed
Hubble parameter and the compactification scale is ten orders less:
\begin{equation}
\frac{H}{\mu_c}\approx\frac{H}{b}\approx 10^{-5}. \label{Hb}
\end{equation}

Of course this geometric reformulation of the hierarchy problem rises the question about the
stability of the brane separation, an issue that must be treated in a similar way to the
Golberger--Wise mechanism \cite{GW} by adding a scalar degree of freedom for this quantity.

\section{Localization of various matter fields on a pure de Sitter thick brane}
\label{SecLocalization}

In this section we shall investigate the localization of various
bulk matter fields on a pure de Sitter thick brane. Spin--0 scalars,
spin--1 vectors and spin--1/2 fermions will be considered by means
of gravitational interaction. Certainly, it has been implicitly
assumed that the various bulk matter fields considered below make
little contribution to the bulk energy so that the solutions given
in the previous section remain valid even in the presence of bulk
matter. Thus, these bulk matter fields make little contribution to
geometry of the bulk spacetime. The mass spectra of various matter
fields on the pure de Sitter thick brane will also be discussed by
presenting and analyzing the potential of the corresponding
Schr\"{o}dinger equation for their KK modes.

In the following, the localization of various bulk matter fields
will be investigated, and it will be seen that the mass-independent
potentials of the corresponding Schr\"{o}dinger equations can be
obtained conveniently with the aid of the metric (\ref{metric_z}).

\subsection{Spin--0 scalar fields}

We begin by considering the localization of real scalar fields (both massless and massive) on the
thick brane obtained in the previous section, turning to vectors and fermions in the next
subsections. Let us start by considering the action of a massive real scalar field coupled to
gravity
\begin{eqnarray}\label{action_scalar}
S_{0}=-\frac{1}{2}\int d^{5}x\sqrt{-g}~
          \left(g^{MN}\partial_{M}\Phi\partial_{N}\Phi+m_s^2\Phi^2\right),
\end{eqnarray}
where $m_s$ is the mass of the bulk scalar field. Using the conformal metric (\ref{metric_z}), the
equation of motion derived from (\ref{action_scalar}) reads
\begin{eqnarray}\label{EqOfScalar5D}
\frac{1}{\sqrt{-\hat{g}}}\partial_{\mu}\left(\sqrt{-\hat{g}} \hat{g}^{\mu \nu}\partial_{\nu}
\Phi\right) + e^{-3A}\partial_{z} \left(e^{3A}\partial_z\Phi\right) - e^{2A}m_s^2\Phi = 0.
\end{eqnarray}
Then, by using the KK decomposition
 $\Phi(x,z)=\sum_{n}\phi_{n}(x)\chi_{n}(z)e^{-3A/2}$ and demanding that
 $\phi_{n}$ satisfies the 4D massive Klein--Gordon
equation:
\begin{eqnarray}
\label{4dKGEq}
\left[\frac{1}{\sqrt{-\hat{g}}}\partial_{\mu}\left(\sqrt{-\hat{g}}
   \hat{g}^{\mu \nu}\partial_{\nu}\right) -m_{n}^{2} \right]\phi_{n}(x)=0,
\end{eqnarray}
where $m_{n}$ is the 4D mass of the KK excitation of the scalar field, we can
obtain the equation for the scalar KK mode $\chi_{n}(z)$:
\begin{eqnarray}
\left[-\partial^{2}_z+ V_{0}(z)\right]{\chi}_n(z)
  =m_{n}^{2} {\chi}_{n}(z),
  \label{SchEqScalar1}
\end{eqnarray}
which is a Schr\"{o}dinger equation with the effective potential given by
\begin{eqnarray}
  V_0(z)=\frac{3}{2} A'' + \frac{9}{4}A'^{2} + e^{2A}m_s^2. \label{VScalar}
\end{eqnarray}
It is clear that the potential $V_{0}(z)$ defined in (\ref{VScalar}) is completely determined by
the warp factor and the bulk mass of the scalar field, hence, it is a 4D
mass--independent potential.

\begin{figure*}[htb]
\begin{center}
\includegraphics[width=7cm]{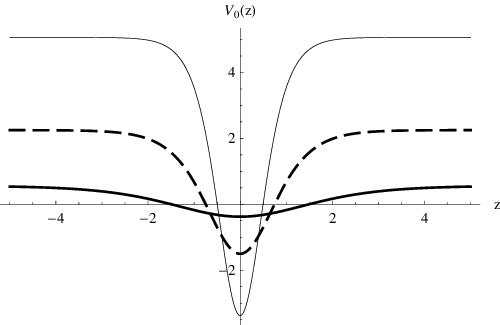}
\end{center}\vskip -5mm
\caption{The shape of the potential of the scalar KK modes $V_{0}(z)$ for different values of $H$,
which is set to $H=0.5$ for the thick line, $H=1.0$ for the dashed line, and $H=1.5$ for the thin
line. Here we have set $m_s=0$ for simplicity.} \label{fig_Scalar_V}
\end{figure*}

The full 5D action (\ref{action_scalar}) can be
reduced to the standard 4D action for a massless and
a series of massive scalars
\begin{eqnarray}
 S_{0}=- \frac{1}{2} \sum_{n}\int d^{4} x \sqrt{-\hat{g}}
     \bigg(\hat{g}^{\mu\nu}\partial_{\mu}\phi_{n}
           \partial_{\nu}\phi_{n}
           +m_{n}^2 \phi^{2}_{n}
     \bigg), \label{ScalarEffectiveAction}
\end{eqnarray}
when integrated over the extra dimension, with the requirement that
Eq. (\ref{SchEqScalar1}) is satisfied and the following
orthonormalization conditions are obeyed:
\begin{eqnarray}
 \int^{\infty}_{-\infty}
 \;\chi_m(z)\chi_n(z) dz=\delta_{mn}.
 \label{normalizationCondition1}
\end{eqnarray}
For the thick brane solution (\ref{Az}), the potential
(\ref{VScalar}) adopts the form
\begin{eqnarray}\label{VScalar1}
V_{0}(z)=\frac{9}{4}H^{2} - \left(\frac{15}{4}-\frac{m_s^2}{b^2}\right)H^{2}\text{sech}^{2}(Hz),
\end{eqnarray}
where the quantity
\begin{eqnarray}\label{beta}
\beta\equiv\frac{15}{4}-\frac{m_s^2}{b^2}
\end{eqnarray}
must be positive definite in order to ensure localization of the massive scalar field. This
potential has a minimum (negative value) equal to
$\left(-\frac{3}{2}+\frac{m_s^{2}}{b^2}\right)H^{2}$ at $z=0$ and a maximum (positive value) equal
to $\frac{9H^{2}}{4}$ at $z=\pm\infty$, ensuring the existence of a mass gap in the spectrum (see
Fig. \ref{fig_Scalar_V}). Then, by inserting (\ref{VScalar1}) into Eq. (\ref{SchEqScalar1}), the
latter turns into the well-known Schr\"{o}dinger equation with $E_{n}=m_{n}^{2}-\frac{9H^{2}}{4}$:
\begin{eqnarray}\label{SchEqScalar2}
 \left[-\partial_{z}^{2} - \left(\frac{15}{4}-\frac{m_s^2}{b^2}\right)H^{2}\text{sech}^{2}(Hz)\right]
    \chi_{n}= E_{n}\chi_{n}.
\end{eqnarray}
After performing the change of variable $v=Hz$, Eq. (\ref{SchEqScalar2}) displays a modified
P\"{o}schl--Teller potential in its standard form:
\begin{eqnarray}\label{SchEqScalar3}
 \left[-\partial_{v}^{2} - \beta \text{sech}^{2}(v)\right]
    \chi_{n}= E_{n}\chi_{n},
\end{eqnarray}
where the energy now reads $E_{n}=\frac{m_n^2}{H^2}-\frac{9}{4}$.

For the Schr\"{o}dinger equation with a modified P\"{o}schl--Teller potential, the energy spectrum
of bound states can be computed exactly if $\beta=n(n+1)$, where $[n]$ is the number of bound
states \cite{Gremm_2000,Liu0708,ThickBrane4,LiuJCAP2009,1009.1684,Diaz_1999}. Since this quantity
must be positive definite, it varies in the interval $0<\beta\le\frac{15}{4}$. The maximum value
that $\beta$ can reach is $\frac{15}{4}$ (when the bulk scalar field becomes massless $m_s=0$)
which corresponds to $n=\frac{3}{2}$, a fact that means that there are at most two bound states in
the spectrum of the massless bulk scalar field, corresponding to $n=0$ and $n=1$. However, as far
as the bulk scalar field becomes massive ($m_s\ne 0$) and, in turn, $\beta$ passes the critical
value $\beta=2$ (or, equivalently, $m_s^2=\frac{7}{4}b^2$), which still corresponds to two bound
states in the spectrum of the bulk scalar field, the number of bound states reduces to one: the
massless mode with $m_0=0$ and energy $E_0=-\frac{9}{4}$. This situation continues till $\beta$
approaches a vanishing value, $\beta\rightarrow 0$, which constitutes its lower bound.

Thus, the crucial relation between the parameters $m_s$ and $b$, defined through (\ref{beta}),
determines whether there exist or not excited massive KK fluctuations localized on the 3--brane as
bound states of the mass spectrum of the bulk scalar field. In other words, if $\beta<2$ (i. e.,
$m_s>\frac{\sqrt{7}}{2}b$) there will be just one bound state in the mass spectrum of the scalar
field. Moreover, since $\beta$ must be positive definite, this ensures that there will always
exists a massless mode corresponding to $[n]=0$. On the other hand, when $\beta\le 0$, then the
scalar field KK modes are not localized on the 3--brane. Thus, if there is only one bound state
(the massless one), the continuum of massive fluctuations will be delocalized (having access to
travel along the extra dimension), but it will still be separated from the massless ground state
by the mass gap, defined by the asymptotic value of the potential
$V_0(z=\pm\infty)=\frac{9H^{2}}{4}$.

\begin{figure*}[htb]
\begin{center}
\subfigure[$\chi_{0}(z)$]{\label{fig_Scalar_zero}
\includegraphics[width=7cm]{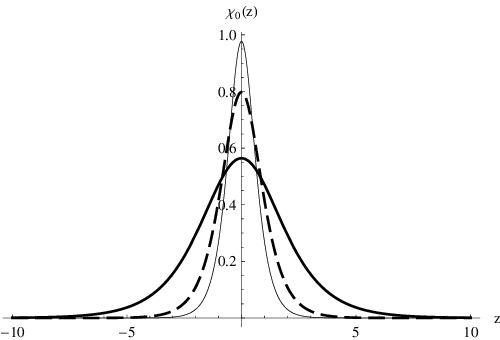}}
\subfigure[$\chi_{1}(z)$]{\label{fig_Scalar_Sec}
\includegraphics[width=7cm]{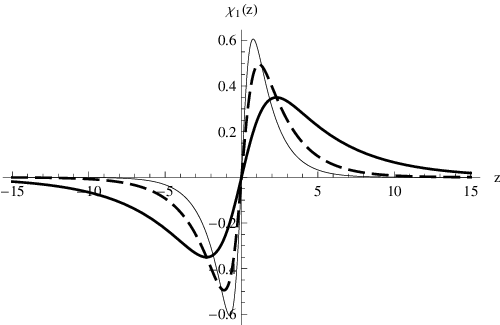}}
\end{center}\vskip -5mm
\caption{The shape of the scalar zero mode $\chi_{0}(z)$ in (a), and the first excited state
$\chi_{1}(z)$ in (b). The parameter $H$ is set to $H=0.5$ for the thick line, $H=1.0$ for the
dashed line, and $H=1.5$ for the thin line. Here we have set $m_s=0$ for simplicity.}
 \label{fig_Scalar_chi}
\end{figure*}

\begin{figure*}[htb]
\begin{center}
\includegraphics[width=7cm]{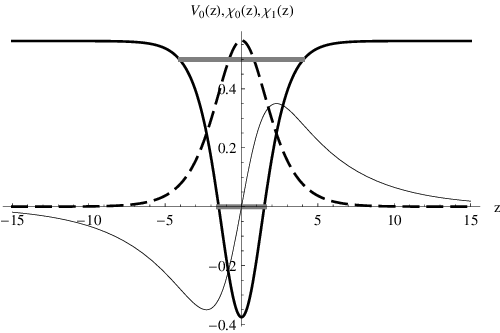}
\end{center}\vskip -5mm
\caption{The shape of the potential of scalar KK modes $V_{0}$ (thick line), the bound scalar KK
modes $\chi_{n}$ (dashed line for $\chi_{0}$ and thin line for $\chi_{1}$) and the mass spectrum
(thick grey lines) with the parameter $H=0.5$. Here we have set $m_s=0$ for simplicity.}
\label{fig_Scalar_all}
\end{figure*}

For the sake of generality we shall come back to Eq. (\ref{SchEqScalar2}), written in terms of the
$z$ coordinate, and consider a massless bulk scalar field ($m_s=0$). In this case the energy
spectrum of bound states reads:
\begin{eqnarray}\label{EnScalar}
E_{n}=-H^{2}\left(\frac{3}{2}-n\right)^{2}
\end{eqnarray}
or, in terms of the 4D squared mass:
\begin{eqnarray}\label{MnScalar}
m_{n}^{2}=n(3-n)H^{2}.
\end{eqnarray}
By recalling that the effective 4D Planck mass (\ref{MPl1}) provides a relation between 4D and 5D
parameters and mass scales of the model, one can alternatively express the 4D energy $E_{n}$ and
mass $m_{n}$ in terms of the following combination of brane and bulk parameters:
\begin{eqnarray}\label{En_MnScalar}
E_{n}=-\frac{2}{\pi}\left(\frac{3}{2}-n\right)^{2}\frac{M_{Pl}^2 b^3}{M_*^3}, \qquad \qquad
m_{n}^{2}=\frac{2n(3-n)}{\pi}\frac{M_{Pl}^2 b^3}{M_*^3}.
\end{eqnarray}
Here $n$ is an integer that satisfies $0\leq n<\frac{3}{2}$, so that it is clear that there are
two bound states. The first one is the ground state with $m_{0}^{2}=0$, and can be expressed as
\begin{eqnarray}\label{ZeroScalar}
\chi_{0}(z)=\sqrt{\frac{2H}{\pi}}\text{sech}^{\frac{3}{2}}(Hz),
\end{eqnarray}
which is just the massless mode and also shows that there is no
tachyonic scalar mode. The second one is the first excited state,
which corresponds to $n=1$ and $m_{1}^{2}=2H^{2}$, and can be
written as
\begin{eqnarray}\label{SecScalar}
\chi_{1}(z)= \sqrt{\frac{2 H}{\pi}} \text{sech}^{\frac{3}{2}}(Hz)\sinh(Hz).
\end{eqnarray}
The shapes of the bound KK modes and the mass spectrum are shown in Fig. \ref{fig_Scalar_chi} and
Fig. \ref{fig_Scalar_all}. The continuous spectrum starts at $m^{2}={9H^{2}}/{4}$. So when the
energy of the scalars is larger than $\frac{3H}{2}$, the scalars cannot be trapped on the brane,
they will be excited into the bulk, which translates into a good chance to experimentally discover
the extra dimensions.

\subsection{Spin--1 vector fields}

We now turn to spin 1 vector fields. We begin with the
5D action of a vector field
\begin{eqnarray}\label{action_Vector}
S_{1} = -\frac{1}{4}\int d^{5}x \sqrt{-g}~ g^{M N}
 g^{RS}F_{MR}F_{NS},
\end{eqnarray}
where $F_{MN}=\partial_{M}A_{N}-\partial_{N}A_{M}$ is the field
strength tensor as usual. From this action the equations of motion
are read as follows
\begin{eqnarray}
\frac{1}{\sqrt{-g}} \partial_{M}\left(\sqrt{-g} g^{M N} g^{R S}
F_{NS}\right) = 0.
\end{eqnarray}
By using the background metric (\ref{metric_z}), the equations of
motion can be written as
\begin{eqnarray}
 \frac{1}{\sqrt{-\hat{g}}}\partial_\nu\left(\sqrt{-\hat{g}} ~
      \hat{g}^{\nu \rho}\hat{g}^{\mu\lambda}F_{\rho\lambda}\right)
    +{\hat{g}^{\mu\lambda}}e^{-A}\partial_z
      \left(e^{A} F_{5\lambda}\right)  = 0, ~~\\
 \partial_\mu\left(\sqrt{-\hat{g}}~ \hat{g}^{\mu \nu} F_{\nu 5}\right) =
 0.~~
\end{eqnarray}

On the other hand, for a 5D vector field, we can make the following general KK decomposition:
\begin{eqnarray}
 A_{M}(x^{\lambda},z)=\sum_{n} a_{M}^{(n)}(x^{\lambda})\rho_{n}(z). \label{KKdecompositionOfAM}
\end{eqnarray}
The action of the 5D massless vector field (\ref{action_Vector}) is invariant under the following
gauge transformation with an arbitrary regular scalar function $F(x^{\lambda},z)$:
\begin{eqnarray}
 A_{M}(x^{\lambda},z) \rightarrow \widetilde{A}_{M}(x^{\lambda},z) &=&
  A_{M}(x^{\lambda},z)+\partial_{M}F(x^{\lambda},z), \label{GaugeTransformation}
\end{eqnarray}
or
\begin{eqnarray}
 A_{\mu}(x^{\lambda},z) \rightarrow \widetilde{A}_{\mu}(x^{\lambda},z) &=&
  A_{\mu}(x^{\lambda},z)+\partial_{\mu}F(x^{\lambda},z), \label{GaugeTransformationA} \\
  A_{5}(x^{\lambda},z) \rightarrow \widetilde{A}_{5}(x^{\lambda},z) &=&
  A_{5}(x^{\lambda},z)+\partial_{z}F(x^{\lambda},z).  \label{GaugeTransformationB}
\end{eqnarray}
Now we will check whether the component $A_{5}(x^{\lambda},z)$ can be set to zero by making use of
the above gauge transformation.

Since we have infinite extra dimensions in our braneworld scenario, i.e. $-\infty < x^{\lambda}, z
< \infty$, we have no any constrain (coming from the topology of the extra dimension) on the gauge
potential $A_M(x^{\lambda},z)$ and the scalar function $F(x^{\lambda},z)$. From the transformation
(\ref{GaugeTransformationB}) and the KK decomposition (\ref{KKdecompositionOfAM}), we have
\begin{eqnarray}
 A_{5}(x^{\lambda},z) \rightarrow \widetilde{A}_{5}(x^{\lambda},z)
   = \sum_{n}a_{5}^{(n)}(x^{\lambda})\rho_{n}(z)+\partial_{z}F(x^{\lambda},z).
\end{eqnarray}
Thus, if we choose the scalar function $F(x^{\lambda},z)$ as
\begin{eqnarray}
  F(x^{\lambda},z)= -\sum_{n}
           a_{5}^{(n)}(x^{\lambda})\int\rho_{n}(z) dz, \label{F(x,z)}
\end{eqnarray}
then the fourth component $\widetilde{A}_5$ will vanish:
\begin{eqnarray}
 \widetilde{A}_{5}(x^{\lambda},z) = 0,
\end{eqnarray}
which is just the gauge choice we made.

\begin{figure*}[htb]
\begin{center}
\includegraphics[width=7cm]{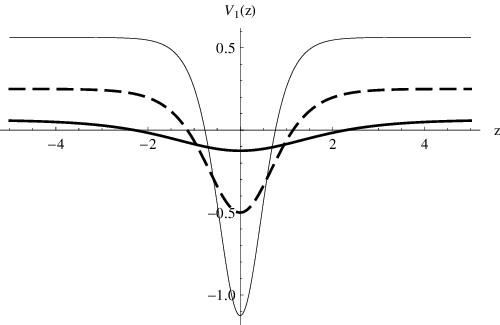}
\end{center}\vskip -5mm
\caption{The shape of the potential of the vector fields $V_{1}(z)$.
The parameter $H$ is set to $H=0.5$ for the thick line, $H=1.0$ for
the dashed line, and $H=1.5$ for the thin line. }
\label{fig_Vector_V}
\end{figure*}

\begin{figure*}[htb]
\begin{center}
\includegraphics[width=7cm]{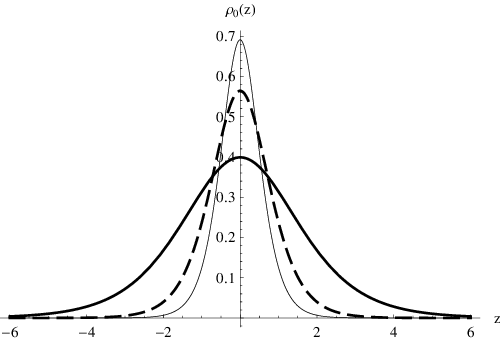}
\end{center}\vskip -5mm
\caption{The shape of the vector zero mode $\rho_{0}(z)$. The
parameter $H$ is set to $H=0.5$ for the thick line, $H=1.0$ for the
dashed line, and $H=1.5$ for the thin line.} \label{fig_Vector_rho}
\end{figure*}

Thus, we choose $A_5=0$ by using this gauge freedom. Then, the action (\ref{action_Vector}) can be
reduced to
\begin{eqnarray}
S_1 = - \frac{1}{4} \int d^5 x \sqrt{-g} \bigg\{
        g^{\mu\alpha} g^{\nu\beta} F_{\mu\nu}F_{\alpha\beta}
        +2e^{-A} g^{\mu\nu} \partial_z A_{\mu} \partial_z A_{\nu}
       \bigg\}.
\label{actionVector2}
\end{eqnarray}
With the decomposition of the vector field $A_{\mu}(x,z)=\sum_n
a^{(n)}_\mu(x)\rho_n(z)e^{-A/2}$ and the orthonormalization
conditions
\begin{eqnarray}
 \int^{\infty}_{-\infty}  \;\rho_m(z)\rho_n(z)dz=\delta_{mn},
 \label{normalizationCondition2}
\end{eqnarray}
the action (\ref{actionVector2}) is read
\begin{eqnarray}
S_1 = \sum_{n}\int d^4 x \sqrt{-\hat{g}}~
       \bigg( - \frac{1}{4}\hat{g}^{\mu\alpha} \hat{g}^{\nu\beta}
             f^{(n)}_{\mu\nu}f^{(n)}_{\alpha\beta}
       - \frac{1}{2}m_{n}^2 ~\hat{g}^{\mu\nu}
           a^{(n)}_{\mu}a^{(n)}_{\nu}
       \bigg),
\label{actionVector3}
\end{eqnarray}
where $f^{(n)}_{\mu\nu} = \partial_\mu a^{(n)}_\nu - \partial_\nu
a^{(n)}_\mu$ is the 4D field strength tensor. In the
above process, it has been required that the vector KK modes
$\rho_n(z)$ should satisfy the following Schr\"{o}dinger equation:
\begin{eqnarray}
   \left[-\partial^2_z +V_1(z) \right]{\rho}_n(z)=m_n^{2}   {\rho}_n(z),
    \label{SchEqVector1}
\end{eqnarray}
where the mass--independent potential $V_{1}(z)$ is given by
\begin{eqnarray}
\label{V_Vector}
 V_{1}(z)=\frac{H^{2}}{4}-\frac{3H^{2}}{4}\text{sech}^{2}(Hz).
\end{eqnarray}
The potential also has the a minimum $-{H^{2}}/{2}$ at $z=0$ and a maximum ${H^{2}}/{4}$ at $z=\pm
\infty$ (see Fig. \ref{fig_Vector_V}), a fact that ensures the presence of a mass gap in the spectrum
of KK vector modes. Eq. (\ref{SchEqVector1}) with this potential turns into the
following Schr\"{o}dinger equation with a modified P\"{o}schl-Teller potential:
\begin{eqnarray}\label{SchEqVector2}
\left[-\partial_{z}^{2}-\frac{3H^{2}}{4}\text{sech}^{2}(Hz)\right]\rho_{n}=E_{n}\rho_{n},
\end{eqnarray}
where $E_{n}=m_{n}^{2}-\frac{H^{2}}{4}$. The energy spectrum of
bound states can be expressed as follows:
\begin{eqnarray}\label{EnVector}
E_{n}=-H^{2}\left(\frac{1}{2}-n\right)^{2}
\end{eqnarray}
or, in terms of the squared mass:
\begin{eqnarray}\label{MnVector}
m_{n}^{2}=n(1-n)H^{2}.
\end{eqnarray}
Here $n$ is an integer satisfying $0\leq n <\frac{1}{2}$. So
there is only one bound state (the ground state), i.e., the normalized
zero mode
\begin{eqnarray}\label{ZeroVector}
\rho_{0}(z)=\sqrt{\frac{H}{\pi}\text{sech}(Hz)}
\end{eqnarray}
with $m_{0}^{2}=0$. The shape of the zero mode is shown in Fig. \ref{fig_Vector_rho}. 
On the other hand, if we further perform the following rescaling of the fifth coordinate
$u=Hz$ in order to write Eq. (\ref{SchEqVector2}) in the canonical form
\begin{eqnarray}\label{SchEqVector3}
\left[-\partial_{u}^{2}-\frac{3}{4}\text{sech}^{2}(u)\right]\rho_{n}(u)=E_{n}\rho_{n}(u),
\end{eqnarray}
where now $E_{n}=\frac{m_{n}^{2}}{H^{2}}-\frac{1}{4}$, then we can express the general  
solution for the continuum of massive KK vector modes as
\begin{equation}
\rho_n(z) = C_1(\beta) P^\mu_{1/2}\left(\tanh u\right) + C_2(\beta) Q^\mu_{1/2}\left(\tanh u\right),
\label{gralsolnPQ}
\end{equation}
where $C_1(\beta)$ and $C_2(\beta)$ are arbitrary constants which depend on $\beta$; $P^\mu_{1/2}$ and $Q^\mu_{1/2}$ are
associated Legendre functions of first and second kind, respectively, degree $\nu=1/2$ and purely 
imaginary order $\mu=i\beta=i\sqrt{m^2/H^2-1/4}$. These functions are linearly independent.
It is worth mentioning that due to the following relation between the associated Legendre 
functions
\begin{equation}
Q^\mu_{\nu}\left(x\right) = \frac{\pi}{2\sin(\pi\mu)}\left[
P^\mu_{\nu}\left(x\right)\cos(\pi\mu) - \frac{\Gamma(\nu+\mu+1)}{\Gamma(\nu-\mu+1)} P^{-\mu}_{\nu}\left(x\right)\right],
\label{relnPQ}
\end{equation}
the expression for the massive KK vector modes can alternatively be given in terms of the following
pair of independent associated Legendre functions (in the language of the conformal variable $z$)
\begin{equation}
\rho_n(z) = C_+(\beta) P^{+i\beta}_{1/2}\left(\tanh(Hz)\right) + C_-(\beta) P^{-i\beta}_{1/2}\left(\tanh(Hz)\right),
\label{gralsolnPP}
\end{equation}
which are real functions and asymptotically behave as plane waves for $z\rightarrow\infty$
\begin{equation}
P_{1/2}^{\pm i\beta}\left(\text{tanh}(Hz)\right)\sim \frac{e^{\pm i\beta Hz}}{\Gamma\left(1\mp i\beta\right)}.
\label{Pwaves}
\end{equation}
This behaviour leads to the following values for the constants $C_+(\beta)$ and $C_-(\beta)$ when normalizing the 
wave functions in the plane wave sense, i.e. in a box of length $2\pi$: 
\begin{equation}
C_{\pm}(\beta) = \frac{\Gamma\left(1\mp i\beta\right)}{\sqrt{2\pi}} = \frac{\left|\Gamma\left(1+i\beta\right)\right|}{\sqrt{2\pi}}
\label{C+C-}
\end{equation}
since $\left|\Gamma\left(1+i\beta\right)\right|=\left|\Gamma\left(1-i\beta\right)\right|$ and we need just the module of these 
Gamma functions. These expressions will be 
very useful later on, when computing the corrections to Coulomb law coming from the massive extra dimensional modes.

\begin{figure*}[htb]
\begin{center}
\includegraphics[width=7cm]{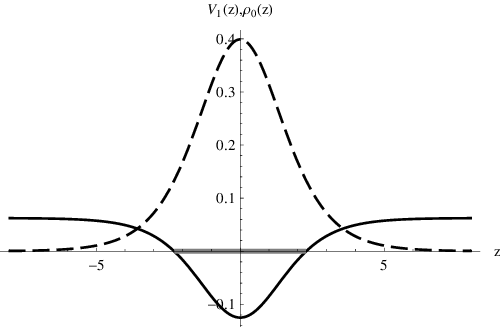}
\end{center}\vskip -5mm
\caption{The shape of the potential of vector KK modes $V_{1}$
(thick line), the vector zero mode $\rho_{0}$ (dashed line), and the
mass spectrum (thick grey lines) with the parameter $H=0.5$. }
\label{fig_Vector_all}
\end{figure*}

It is easy to see that there is a mass gap between the zero mode and the excited modes from Fig.
\ref{fig_Vector_all} since the continuous spectrum of massive vector KK modes starts at
$m^{2}={H^{2}}/{4}$ and asymptotically turns into plane waves in agreement with (\ref{Pwaves}). Similar to the case 
of scalars, when the energy of vectors is larger than $\frac{1}{2}H$, they are not localized on the brane anymore,
leaking to the extra dimension.

\subsection{Spin--1/2 fermion fields}

Finally, we will study the localization of fermions on the pure de Sitter thick braneworld. In
5D spacetime, fermions are four--component spinors and their Dirac structure can be
described by $\Gamma^M= e^M_{~\bar{M}} \Gamma^{\bar{M}}$ with $e^M_{~\bar{M}}$ being the vielbein
and $\{\Gamma^M,\Gamma^N\}=2g^{MN}$. In this Subsection, $\bar{M}, \bar{N}, \cdots =0,1,2,3,5$ and
$\bar{\mu}, \bar{\nu}, \cdots =0,1,2,3$ denote the 5D and 4D local
Lorentz indices, respectively, and $\Gamma^{\bar{M}}$ are the gamma matrices in 5D
flat spacetime. In our set-up, the vielbein is given by
\begin{eqnarray}
e_M ^{~~\bar{M}}= \left(%
\begin{array}{ccc}
  \text{e}^{A} \hat{e}_\mu^{~\bar{\nu}} & 0  \\
  0 & \text{e}^{A}  \\
\end{array}%
\right),\label{vielbein_e}
\end{eqnarray}
$\Gamma^M=\text{e}^{-A}(\hat{e}^{\mu}_{~\bar{\nu}}
\gamma^{\bar{\nu}},\gamma^5)=\text{e}^{-A}(\gamma^{\mu},\gamma^5)$,
where $\gamma^{\mu}=\hat{e}^{\mu}_{~\bar{\nu}}\gamma^{\bar{\nu}}$,
$\gamma^{\bar{\nu}}$ and $\gamma^5$ are the usual flat gamma
matrices in the 4D Dirac representation. The Dirac
action of a spin--1/2 fermion with a mass term can be expressed as
\cite{PRD_Oda_026002}
\begin{eqnarray}
S_{\frac{1}{2}} = \int d^5 x \sqrt{-g} \left[\bar{\Psi} \Gamma^M
          \left(\partial_M+\omega_M\right) \Psi
          - M F(z) \bar{\Psi}\Psi\right]. \label{DiracAction}
\end{eqnarray}
Here $\omega_M$ is the spin connection defined as $\omega_M=
\frac{1}{4} \omega_M^{\bar{M} \bar{N}} \Gamma_{\bar{M}}
\Gamma_{\bar{N}}$ with
\begin{eqnarray}
 \omega_M ^{\bar{M} \bar{N}}
   &=& \frac{1}{2} {e}^{N \bar{M}}\left(\partial_M e_N^{~\bar{N}}
                      - \partial_N e_M^{~\bar{N}}\right)
    - \frac{1}{2} {e}^{N\bar{N}}\left(\partial_M e_N^{~\bar{M}}
                      - \partial_N e_M^{~\bar{M}}\right)  \nonumber \\
   && - \frac{1}{2} {e}^{P \bar{M}} {e}^{Q \bar{N}}\left(\partial_P e_{Q
{\bar{R}}} - \partial_Q e_{P {\bar{R}}}\right) {e}_M^{~\bar{R}},
\end{eqnarray}
and $F(z)$ is some general scalar function of the extra dimensional
coordinate $z$. We will discuss about the properties of the scalar
function $F(z)$ later on, in the context of the localization of KK
fermion modes. The non--vanishing components of the spin connection
$\omega_M$ for the background metric (\ref{metric_z}) are
\begin{eqnarray}
  \omega_\mu =\frac{1}{2}(\partial_{z}A) \gamma_\mu \gamma_5
             +\hat{\omega}_\mu, \label{spinConnection}
\end{eqnarray}
where $\hat{\omega}_\mu=\frac{1}{4} \bar\omega_\mu^{\bar{\mu}
\bar{\nu}} \Gamma_{\bar{\mu}} \Gamma_{\bar{\nu}}$ is the spin
connection derived from the metric
$\hat{g}_{\mu\nu}(x)=\hat{e}_{\mu}^{~\bar{\mu}}(x)
\hat{e}_{\nu}^{~\bar{\nu}}(x)\eta_{\bar{\mu}\bar{\nu}}$. Thus, the
equation of motion corresponding to the action (\ref{DiracAction})
can be written as
\begin{eqnarray}
 \left[ \gamma^{\mu}(\partial_{\mu}+\hat{\omega}_\mu)
         + \gamma^5 \left(\partial_z  +2 \partial_{z} A \right)
         -\text{e}^A MF(z)
 \right ] \Psi =0, \label{DiracEq1}
\end{eqnarray}
where $\gamma^{\mu}(\partial_{\mu}+\hat{\omega}_\mu)$ is the Dirac
operator on the brane.

Next, we will investigate the 5D Dirac equation
(\ref{DiracEq1}), and write the spinor in terms of 4D
effective fields. On account of the fifth gamma matrix $\gamma^{5}$,
we anticipate the left-- and right--handed projections of the
4D part to behave differently. From Eq.
(\ref{DiracEq1}), the solutions of the general chiral decomposition
is found to be
\begin{equation}
 \Psi= \text{e}^{-2A}\left(\sum_n\psi_{Ln}(x) L_n(z)
 +\sum_n\psi_{Rn}(x) R_n(z)\right),
\end{equation}
where $\psi_{Ln}(x)=-\gamma^5 \psi_{Ln}(x)$ and
$\psi_{Rn}(x)=\gamma^5 \psi_{Rn}(x)$ are the left-handed and
right-handed components of a 4D Dirac field,
respectively. Hence, we assume that $\psi_{Ln}(x)$ and
$\psi_{Rn}(x)$ satisfy the 4D Dirac equations. Then
the KK modes $L_{n}(z)$ and $R_{n}(z)$ should satisfy the following
coupled equations:
\begin{subequations}\label{CoupleEq1}
\begin{eqnarray}
 \left[\partial_z
                  + \text{e}^A MF(z) \right]L_n(z)
  &=&  ~~m_n R_n(z), \label{CoupleEq1a}  \\
 \left[\partial_z
                  - \text{e}^A MF(z) \right]R_n(z)
  &=&  - m_n L_n(z). \label{CoupleEq1b}
\end{eqnarray}
\end{subequations}
From the above coupled equations, we can obtain the
Schr\"{o}dinger--like equations for the left-- and right--chiral KK
modes of fermions:
\begin{eqnarray}
  \big(-\partial^2_z + V_L(z) \big)L_n
            &=&m_{L_n}^{2} L_n,~~
   \label{SchEqLeftFermion}  \\
  \big(-\partial^2_z + V_R(z) \big)R_n
            &=&m_{R_n}^{2} R_n,
   \label{SchEqRightFermion}
\end{eqnarray}
where the mass--independent potentials are given by
\begin{subequations}\label{Vfermion}
\begin{eqnarray}
  V_L(z)&=& \text{e}^{2A} M^{2}F^{2}(z)
     - \text{e}^{A} A' M F(z) -\text{e}^{A}M\partial_{z}F(z), \label{VL}\\
  V_R(z)&=&   \text{e}^{2A} M^{2}F^{2}(z)
     + \text{e}^{A} A' M F(z) + \text{e}^{A}M\partial_{z}F(z). \label{VR}
\end{eqnarray}
\end{subequations}

For the purpose of getting the standard 4D action for
a massless fermion and a series of massive chiral fermions
\begin{eqnarray}
 S_{\frac{1}{2}} &=& \int d^5 x \sqrt{-g} ~\bar{\Psi}
     \left[ \Gamma^M (\partial_M+\omega_M)
     -MF(z)\right] \Psi  \nonumber \\
  &=&\sum_{n}\int d^4 x \sqrt{-\hat{g}}
    ~\bar{\psi}_{n}
      \left[\gamma^{\mu}(\partial_{\mu}+\hat{\omega}_\mu)
        -m_{n}\right]\psi_{n},~~~
\end{eqnarray}
the following orthonormalization conditions for $L_{n}$ and $R_{n}$
are needed:
\begin{eqnarray}
 \int_{-\infty}^{+\infty} L_m L_ndz
   &=& \delta_{mn}, \label{orthonormalityFermionL} \\
 \int_{-\infty}^{+\infty} R_m R_ndz
   &=& \delta_{mn}, \label{orthonormalityFermionR}\\
 \int_{-\infty}^{+\infty} L_m R_ndz
   &=& 0. \label{orthonormalityFermionLR}
\end{eqnarray}

If in the formulae (\ref{CoupleEq1a}) and (\ref{CoupleEq1b}) one
sets $m_n=0,$ then one gets
\begin{subequations}
\begin{eqnarray}
  L_0&\propto & e^{-M \int e^{A} F dz}, \label{zerol}\\
  R_0&\propto &  e^{M \int e^{A} F dz}. \label{zeror}
\end{eqnarray}
\end{subequations}
The above relations tell us that it is not possible to have both massless left- and right-chiral
KK fermion modes localized on the brane at the same time, since when one is normalizable, the
other one is not.

From Eqs. (\ref{SchEqLeftFermion}), (\ref{SchEqRightFermion}) and
(\ref{Vfermion}), we can see that, if we do not introduce the mass
term in the action (\ref{DiracAction}), i.e., when $M=0$, the
potentials for left-- and right--chiral KK modes $V_{L,R}(z)$ will
vanish and both left-- and right--chiral fermions cannot be
localized on the thick brane. Moreover, if we demand $V_{L}(z)$ and
$V_{R}(z)$ to be $Z_{2}$-even with respect to the extra dimension
$z$, then the mass term $MF(z)$ must be an odd function of $z$. In
this paper, we will consider four cases: $F(z)=\varepsilon(z)$,
$F(z)=\tanh(Hz)$, $F(z)=\sinh(Hz)$ and
$F(z)=(Hz)^{2k+1}\text{e}^{-A}.$

\subsubsection{Case I: $F(z)=\varepsilon(z)$}

\begin{figure*}[htb]
\begin{center}
\includegraphics[width=7cm]{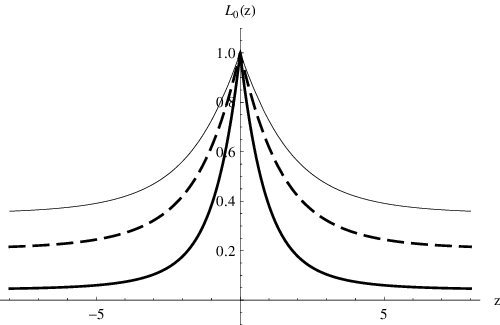}
\end{center}\vskip -5mm
\caption{The shape of the zero mode of left--chiral fermions
$L_{0}(z)$ for different values of the parameter $b$. The
parameters are set to $M=1.0$, $H=0.5$, and $b=0.5$ for the thick
line, $b=1.0$ for the dashed line, and $b=1.5$ for the thin line.}
\label{fig_Fermion_zero}
\end{figure*}

We shall first consider the simplest case $F(z)=\varepsilon(z)$
\cite{PRD_Oda_026002}, where $\varepsilon(z\neq 0)\equiv \frac{z}{|z|}$
and $\epsilon(0)=0$. The explicit forms of the potentials
(\ref{Vfermion}) can be expressed as follows:
\begin{subequations}\label{VfermionCaseI}
\begin{eqnarray}
\label{VfermionCaseILeft}
  V_L(z)
        &=& \frac{H^{2}M\text{sech}^{2}(Hz)}{b^{2}}
            \left[M+b\sinh(Hz)\varepsilon(z)\right]
             -2\delta(z),  \\
\label{VfermionCaseIRight}
  V_R(z)
   &=& \frac{H^{2}M\text{sech}^{2}(Hz)}{b^{2}}
            \left[M-b\sinh(Hz)\varepsilon(z)\right]
              +2\delta(z).
\end{eqnarray}
\end{subequations}
From these expressions, we find that the potentials $V_{L,R}(z)$
have the asymptotic behavior:
$V_{L,R}(z\rightarrow\pm\infty)\rightarrow 0$. Because there is a
$\delta$ function in the expressions for both potentials, when
$z=0$, $V_{L}(0)=-\infty$ and $V_{R}(0)=+\infty$. It is clear that
only the potential $V_{L}(z)$ is of volcano type and has a
$\delta$--potential well at the location of the brane, so just the
massless mode of left-chiral fermion might be trapped on the brane.
The left-chiral fermion zero mode can be computed by solving
(\ref{CoupleEq1a}) with $m_{0}=0$:
\begin{eqnarray}\label{zerofermionI}
L_0(z)
  &\propto&  \exp\left(-\int^z_0 dz'\text{e}^{A(z')} M \epsilon(z')\right).
                   \nonumber\\
  &=&  \exp\left( -\frac{2M}{b} \arctan\left[\tanh\left(\frac{H|z|}{2}\right)\right]\right).
  \label{zeroMode1}
\end{eqnarray}
However, from this expression, we know that when far away from the
brane, $L_{0}(z\rightarrow\pm\infty)$ approaches a positive constant
$\text{e}^{-\frac{\pi M}{2b}}$, a fact which results in the
non--normalization of $L_0(z)$. So the zero mode of left--chiral
fermion cannot be localized on the brane. The shape of the zero mode
$L_0(z)$ is shown in Fig. \ref{fig_Fermion_zero}. Hence, both left-
and right-chiral fermion zero modes cannot be trapped on the brane.
Thus, there is no mass gap in the spectrum of KK modes for both
left-- and right--chiral fermions, the spectrum is continuous and
starts at $m^2=0$.

\subsubsection{Case II: $F(z)=\tanh(Hz)$}

\begin{figure*}[htb]
\begin{center}
\subfigure[$V_{L}(z)$]{\label{fig_fermionII_VL}
\includegraphics[width=7cm]{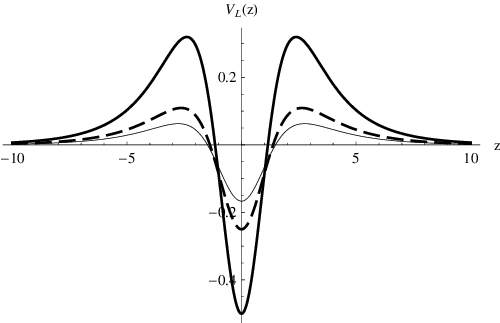}}
\subfigure[$V_{R}(z)$]{\label{fig_fermionII_VR}
\includegraphics[width=7cm]{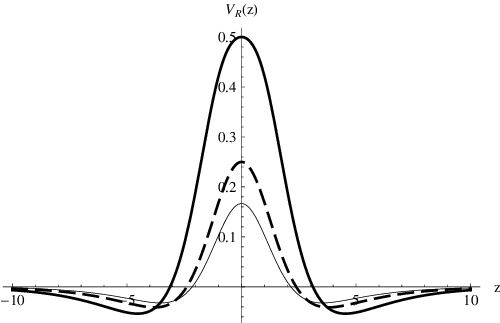}}
\end{center}\vskip -5mm
\caption{The shape of the potential for the left--chiral fermions
$V_{L}(z)$ is displayed in (a), and for the right--chiral fermions
$V_{R}(z)$ is plotted in (b) for the case II. The parameters are set
to $H=0.5$, $M=1$, $b=0.5$ for the thick lines, $b=1.0$ for the
dashed lines and $b=1.5$ for the thin lines. }
 \label{fig_FermionII_V}
\end{figure*}

\begin{figure*}[htb]
\begin{center}
\includegraphics[width=7cm]{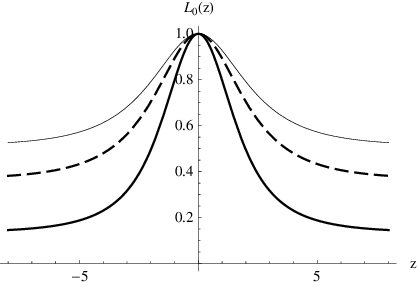}
\end{center}\vskip -5mm
\caption{The shape of the massless left--chiral fermion $\L_{0}(z)$
for case II. The parameters are set to $H=0.5$, $M=1$, $b=0.5$ for
the thick line, $b=1.0$ for the dashed line and $b=1.5$ for the thin
line. } \label{fig_FermionII_zero}
\end{figure*}

For the case $F(z)=\tanh(Hz)$, the explicit forms of the potentials (\ref{Vfermion}) are
\begin{eqnarray}\label{VfermionCaseII}
V_{L}(z)&=& \frac{H^{2}M}{4b^{2}}\text{sech}^{4}(Hz)
          \left[4M\sinh^{2}(Hz)+b\cosh(3Hz)-5b\cosh(Hz)\right], \\
V_{R}(z)&=& \frac{H^{2}M}{4b^{2}}\text{sech}^{4}(Hz)
          \left[4M\sinh^{2}(Hz)-b\cosh(3Hz)+5b\cosh(Hz)\right].
\end{eqnarray}
It can be seen that the potentials $V_{L,R}$ have the following
asymptotic behaviors: they tend to zero as $z\rightarrow\pm\infty$,
and at $z=0$, the potential $V_{L}$ reaches its minimum (negative
value) $-{H^{2}M}/{b}$ while the potential $V_{R}$ has its maximum
(positive value) ${H^{2}M}/{b}$ (see Fig. \ref{fig_FermionII_V}). So
$V_{L}(z)$ is a modified volcano--type potential. For this type of
potentials, there is no mass gap to separate the fermion zero mode
from the excited KK massive modes. Both left-- and right--chiral KK
modes have a continuous gapless spectrum. Because only the potential
for left--chiral fermions has a negative value at the location of
the brane, we only need to study whether the zero mode of
left--chiral fermions $L_0$ could be localized on the brane. The
expression for $L_0$ is
\begin{eqnarray}\label{zerofermionII}
L_{0}\propto \exp\left[\frac{M}{b}(\text{sech(Hz)}-1)\right].
\end{eqnarray}
From this expression and Fig. \ref{fig_FermionII_zero}, we can see
that the zero mode $L_{0}\rightarrow \text{e}^{-\frac{M}{b}}>0$ as
$z\rightarrow\pm\infty$, which indicates that the normalization
condition $\int_{-\infty}^{+\infty}L_{0}^{2}(z)dz<\infty$ is not
satisfied, so the zero mode of the left--chiral fermions cannot be
localized on the brane.

\subsubsection{Case III: $F(z)=\sinh(Hz)$}

\begin{figure*}[htb]
\begin{center}
\subfigure[$V_{L}(z)$]{\label{fig_fermionIII_VL}
\includegraphics[width=7cm]{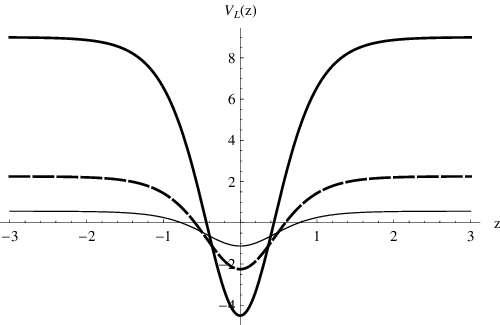}}
\subfigure[$V_{R}(z)$]{\label{fig_fermionIII_VR}
\includegraphics[width=7cm]{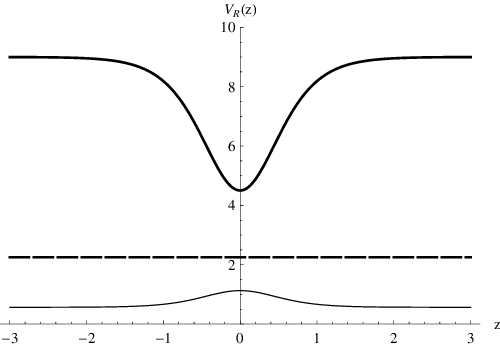}}
\end{center}\vskip -5mm
\caption{The shape of the potential $V_{L}(z)$ is presented in (a)
and $V_{R}(z)$ is displayed in (b) for the case III. The parameters
are set to $H=1.5$, $M=1$, $b=0.5$ for thick lines, $b=1.0$ for
dashed lines and $b=1.5$ for thin lines.}
 \label{fig_FermionIII_V}
\end{figure*}

\begin{figure*}[htb]
\begin{center}
\includegraphics[width=7cm]{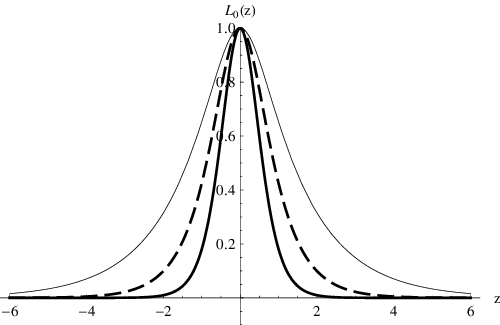}
\end{center}\vskip -5mm
\caption{The shape of the zero mode of the left--chiral fermion
$L_{0}(z)$ for the case III. The parameters are set to $H=1.5$,
$M=1$, $b=0.5$ for the thick line, $b=1.0$ for the dashed line and
$b=1.5$ for the thin line.} \label{fig_FermionIII_zero}
\end{figure*}

For the choice $F(z)=\sinh(Hz)$, the potentials
(\ref{Vfermion}) can be expressed as follows:
\begin{eqnarray}
\label{VLCaseIII}
V_{L}(z)=\frac{H^{2}M}{b^{2}}\left[M-(b+M)\text{sech}^{2}(Hz)\right], \\
\label{VRCaseIII}
V_{R}(z)=\frac{H^{2}M}{b^{2}}\left[M+(b-M)\text{sech}^{2}(Hz)\right].
\end{eqnarray}
From Fig. \ref{fig_FermionIII_V}, we can see that both potentials
have the same asymptotic behavior:
 $V_{L,R}\rightarrow\frac{H^{2}M^{2}}{b^{2}}$ when
$z\rightarrow\pm\infty$. At $z=0$, the potential of left--chiral
fermions has a minimum with negative value $-\frac{H^{2}M}{b}$,
however, the potential of right--chiral fermions has a positive
value one: $\frac{H^{2}M}{b}$ if and only if $M>b$. Thus, just the
potential of left--chiral fermions possesses a negative value at the
location of the brane, so only the zero mode of left--chiral
fermions $L_0$ will be localized on the brane. The solution for
$L_0$ can be written as
\begin{eqnarray}\label{zerofermionIII}
L_{0}(z) =\left[\frac{H~\Gamma\left(\frac{b+2M}{2b}\right)}
{\sqrt{\pi}~\Gamma\left(\frac{M}{b}\right)}\right]^{\frac{1}{2}}
\text{sech}^{\frac{M}{b}}(Hz).
\end{eqnarray}
Because the parameter $M$ and $b$ are positive, the zero mode
$L_{0}$ vanishes at $z\rightarrow\pm\infty$, and it satisfies the
normalization condition. So the massless mode of left--chiral
fermions is localized on the brane. We plot the shape of the
left--chiral fermion zero mode in Fig. \ref{fig_FermionIII_zero}.
However, the zero mode of right--chiral fermions does not exist, a
fact which can be seen from the potential (\ref{VRCaseIII}) and Fig.
\ref{fig_fermionIII_VR}.

We can see that for $m_{L_{n}}^{2}>\frac{H^{2}M^{2}}{b^{2}}$, the
left-chiral fermions can be described by asymptotic plane waves. \

The general bound KK modes $L_n$ for the potential (\ref{VLCaseIII})
can be found to be
\begin{eqnarray}
\label{LnCaseIIIEven}
 L_{n}&\propto& \cosh^{1+\frac{M}{b}}(Hz)~_{2}\text{F}_{1}
  \left(a_{n},b_{n};\frac{1}{2};-\sinh^{2}(Hz)\right)
\end{eqnarray}
for even $n$ and
\begin{eqnarray}
\label{LnCaseIIIOdd}
 L_{n}&\propto& \cosh^{1+\frac{M}{b}}(Hz)\sinh(Hz)~_{2}\text{F}_{1}
  \left(a_{n}+\frac{1}{2},b_{n}+\frac{1}{2};\frac{3}{2};-\sinh^{2}(Hz)\right)
\end{eqnarray}
for odd $n$, where $_{2}\text{F}_{1}$ is the hypergeometric
function, and the parameters $a_{n}$ and $b_{n}$ are given by
\begin{eqnarray}
 a_{n}=\frac{1}{2}(n+1), ~~~~~~~~
 b_{n}=\frac{M}{b}-\frac{1}{2}(n-1).
\end{eqnarray}
The corresponding mass spectrum of the bound states is
\begin{eqnarray}
\label{MnLnCaseIII}
 m_{L_{n}}^{2}=H^{2}\left(\frac{2M}{b} - n\right)n,
 ~~~~~\left(n=0,1,2,\cdots <\frac{M}{b} \right).
\end{eqnarray}
It can be seen that the ground state always belongs to the spectrum of left--chiral KK modes,
which is precisely the zero mode (\ref{zerofermionIII}) with $m_{L_{0}}^{2}=0$. Because the ground
state has the lowest squared mass $m_{L_{0}}^{2}=0$, there are no tachyonic left--chiral fermions.
If $M<b$, there is only one bound state, i.e., the zero mode (\ref{zerofermionIII}). In order to
get massive excited bound states, the condition $M>b$ should be satisfied and their number depends
on the ratio $\frac{M}{b}$ as indicated in (\ref{MnLnCaseIII}).

For the potential of right--chiral fermions $V_{R}(z)$
(\ref{VRCaseIII}), from Fig. \ref{fig_fermionIII_VR}, we can see
that $V_{R}(z)$ is always positive near the location of the brane,
which shows that it cannot trap the zero mode of right--chiral
fermions. For the case $M < b$, the potential $V_{R}$ has a maximum
$\frac{H^{2}M}{b}$ at $z=0$, and has a minimum
$\frac{H^{2}M^{2}}{b^{2}}$ at $z\rightarrow\pm \infty$, i.e.,
$0<V_{R}(z\rightarrow\pm\infty)<V_{R}(z=0)$. So in this case, there
is no any bound state for right-chiral fermions. For the special
case $M=b$, the potential $V_{R}$ is a positive constant:
$V_{R}(z)=\frac{H^{2}M^{2}}{b^{2}}$. Hence, there is still no any
bound state for this case. In the last case $M>b$, we can see that
$0<V_{R}(z=0)<V_{R}(z\rightarrow\pm\infty)$, which indicates that
there exist some bound states, but the ground state is a massive
one:
\begin{eqnarray}
\label{BoundStateRight}
  R_{0}=\left[\frac{H~\Gamma\left(\frac{M}{2b}\right)}
  {\sqrt{\pi}~\Gamma\left(\frac{M-b}{2b}\right)}\right]
         ^{\frac{1}{2}} \cosh^{1-\frac{M}{b}}(Hz),
         \qquad(M>b)
\end{eqnarray}
with the mass determined by
$m_{R_{0}}^{2}=H^{2}\left(\frac{2M}{b}-1\right)>H^{2}>0$. The
general bound states for this case $(M>b)$ are
\begin{eqnarray}
\label{RnCaseIIIEven}
 R_{n}(z)\propto \cosh^{\frac{M}{b}}(Hz)~_{2}\text{F}_{1}
      \left(\frac{1+n}{2},\frac{M}{b}-\frac{1+n}{2};\frac{1}{2};-\sinh^{2}(Hz)\right)
\end{eqnarray}
for even $n$ and
\begin{eqnarray}
\label{RnCaseIIIOdd}
 R_{n}(z)\propto \cosh^{\frac{M}{b}}(Hz)\sinh(Hz) ~_{2}\text{F}_{1}
      \left(1+\frac{n}{2},\frac{M}{b}-\frac{n}{2};\frac{3}{2};-\sinh^{2}(Hz)\right)
\end{eqnarray}
for odd $n$. Then, the corresponding mass spectrum is
\begin{eqnarray}
\label{MnRnCaseIII}
 \!\!\!\!\!\! m_{R_{n}}^{2}= H^{2}\left(\frac{2M}{b}-(n+1)\right)(n+1),
  ~~~\left(M>b,~~n=0,1,2,\cdots <\frac{M}{b}-1 \right).
\end{eqnarray}
By comparing to the mass spectrum of the left--chiral fermions
(\ref{MnLnCaseIII}), we come to the following conclusion: the number
of bound KK modes of right--chiral fermions $N_{R}$ is one less than
that of the left ones $N_{L}$, i.e. $N_{R}=N_{L}-1$.

\begin{figure*}[htb]
\begin{center}
\subfigure[$n=0$]{\label{fig_FermionIII_L0}
\includegraphics[width=5cm]{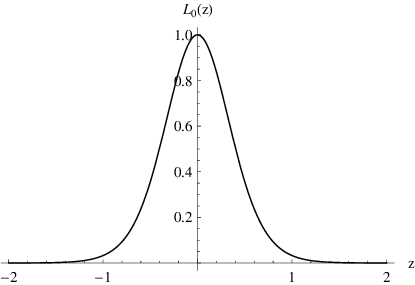}}
\subfigure[$n=1$]{\label{fig_FermionIII_L1}
\includegraphics[width=5cm]{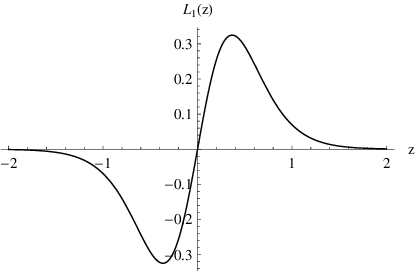}}
\subfigure[$n=2$]{\label{fig_FermionIII_L2}
\includegraphics[width=5cm]{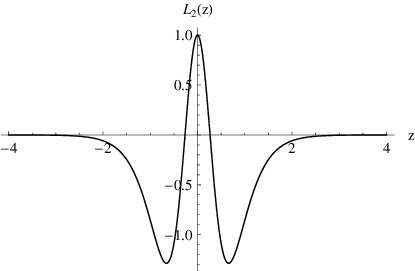}}
\subfigure[$n=3$]{\label{fig_FermionIII_L3}
\includegraphics[width=5cm]{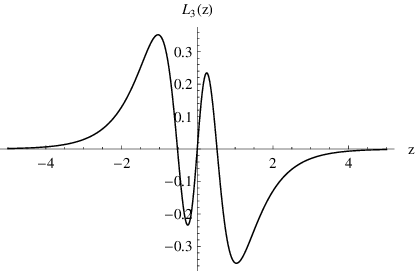}}
\end{center}\vskip -5mm
\caption{The shape of the left--chiral KK modes $L_{n}$ for $0\leq
n\leq 3$ in case III. The parameters are set to $M=1.0$, $b=0.25$
and $H=1.5$.}
 \label{fig_FermionIII_Ln}
\end{figure*}

\begin{figure*}[htb]
\begin{center}
\subfigure[$n=0$]{\label{fig_FermionIII_R0}
\includegraphics[width=4.5cm]{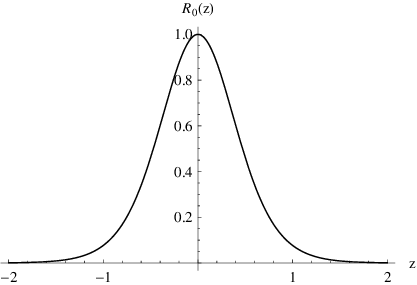}}
\subfigure[$n=1$]{\label{fig_FermionIII_R1}
\includegraphics[width=4.5cm]{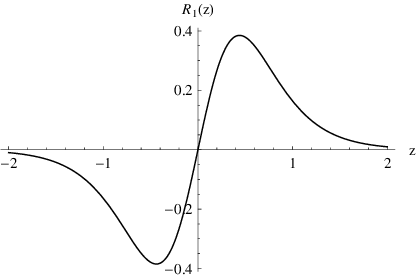}}
\subfigure[$n=2$]{\label{fig_FermionIII_R2}
\includegraphics[width=4.5cm]{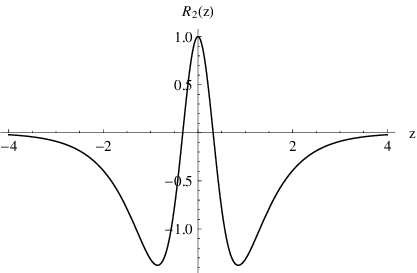}}
\end{center}\vskip -5mm
\caption{The shape of the right--chiral KK modes $R_{n}$ for $0\leq
n\leq 2$ in case III. The parameters are set to $M=1.0$, $b=0.25$
and $H=1.5$.}
 \label{fig_FermionIII_Rn}
\end{figure*}

\begin{figure*}[htb]
\begin{center}
\subfigure[Left chiral fermions]{\label{fig_fermionIII_VLmass}
\includegraphics[width=7cm]{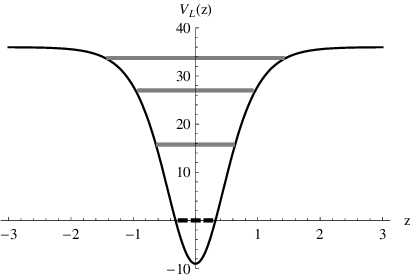}}
\subfigure[Right chiral fermions]{\label{fig_fermionIII_VRmass}
\includegraphics[width=7cm]{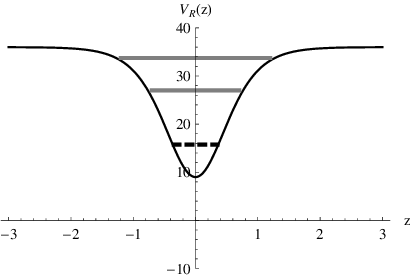}}
\end{center}\vskip -5mm
\caption{The shape of the potentials $V_{L,R}$ (thick lines) and the
mass spectra (the dashed lines correspond to $m^2_{L_0}=0$ and
$m^2_{R_0}=0$, and the grey lines correspond to the bound massive
levels) for the case III. The parameters are set to $M=1.0$,
$b=0.25$ and $H=1.5$.} \label{fig_FermionIII_mass}
\end{figure*}

When $M<b$, there is only one left--chiral fermion bound KK mode (the zero mode), so only the
4D massless left-chiral fermion can be localized on the brane. When $M>b$, there
are $N$ $(N=N_{L}\geq 2)$ bound left--chiral fermion KK modes and $N-1$ $(N_{R}=N-1)$
right--chiral ones. Hence, we obtain that the 4D massless left--chiral fermion and
the massive Dirac fermions consisting of pairs of coupled left-- and right--chiral KK modes can be
localized on the brane. The KK modes $L_{n}$ and $R_n$ are plotted in Fig. \ref{fig_FermionIII_Ln}
and Fig. \ref{fig_FermionIII_Rn}, respectively, and the corresponding mass spectra are shown in
Fig. \ref{fig_FermionIII_mass}. When $m_{L,R}^{2}>\frac{H^{2}M^{2}}{b^{2}}$, both the left-- and
right--chiral fermion KK modes cannot be confined to the brane and will be excited into the bulk.

\subsubsection{Case IV: $F(z)=(Hz)^{2k+1}\text{e}^{-A(z)}$}
Now let us consider the following family of functions:
\begin{equation}
F(z)=(Hz)^{2k+1}\text{e}^{-A(z)},\label{Fclass}
\end{equation}
where the factor $Hz$ renders a dimensionless function $F(z)$ and
$k=0,1,2,\cdots$.

\begin{figure*}[htb]
\begin{center}
\includegraphics[width=7cm]{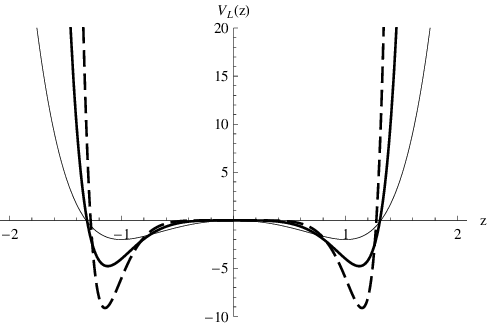}
\end{center}\vskip -5mm
\caption{The shape of the potential $V_{L}$ is represented for
different values of $k:$ $k=1$ (thin line), $k=2$ (thick line) and
$k=3$ (dashed line). In all these cases the potential $V_{L}$ has
two negative minima and becomes infinite as $z\rightarrow\pm\infty$,
resembling a shifted mexican hat potential; thus, there is a
localized massless KK mode and a tower of discrete massive modes.
The parameters are set to $H=1$ and $M=1$ in the figure.}
\label{Vleftgeneral}
\end{figure*}

\begin{figure*}[htb]
\begin{center}
\includegraphics[width=7cm]{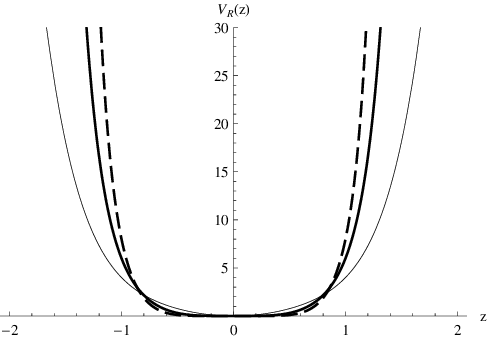}
\end{center}\vskip -5mm
\caption{The shape of the potential $V_{R}$ is represented for
different values of $k:$ $k=1$ (thin line), $k=2$ (thick line) and
$k=3$ (dashed line). In all cases there is a single minimum of the
potential $V_{R}$ at $z=0$ and it is infinite as
$z\rightarrow\pm\infty.$ In this case we do not have a localized
massless zero mode and all massive KK modes are localized on the
brane. Here, we also set $H=1$ and $M=1$.} \label{Vrightgeneral}
\end{figure*}

By using (\ref{warpfactorz}), the explicit form of the potentials
(\ref{Vfermion}) can be expressed as follows
\begin{subequations}
\begin{eqnarray}
V_L(z)&=& MH\left[MH (Hz)^{4k}z^{2}-(2k+1)(Hz)^{2k}\right],\label{leftpotential} \\
V_R(z)&=& MH\left[MH
(Hz)^{4k}z^{2}+(2k+1)(Hz)^{2k}\right].\label{rightpotential}
\end{eqnarray}
\end{subequations}
Both potentials have the same asymptotic behavior $V_{L,R}
\longrightarrow \infty$ when $z\longrightarrow\pm\infty$ and are
bounded below as is shown in Figs. \ref{Vleftgeneral} and
\ref{Vrightgeneral} for different values of $k>0,$ the case $k=0$
will be considered below as a separate one. Then, the KK spectrum is
discrete for both kind of fermions. It is possible to show using
(\ref{zerol}) that the zero left-chiral KK mode has the following
form

\begin{figure*}[htb]
\begin{center}
\includegraphics[width=7cm]{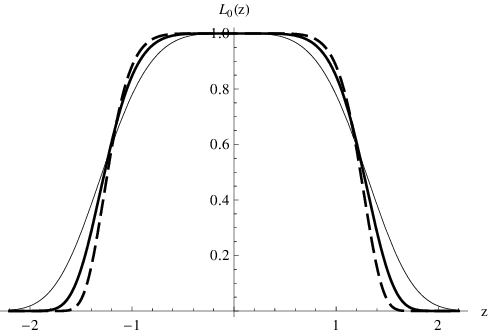}
\end{center}\vskip -5mm
\caption{The shape of the massless left--chiral KK mode $L_{0}(z)$
is represented for different values of $k:$ $k=1$ (thin line), $k=2$
(thick line) and $k=3$ (dashed line). The parameters are set to
$H=1$ and $M=1$.}
 \label{fig_FermionIV_Zero}
\end{figure*}
%
\begin{equation}
L_0 \propto
\text{e}^{-\frac{M}{2H(k+1)}(Hz)^{2(k+1)}}.\label{zeromodeG}
\end{equation}
Thus, the massless left-chiral KK mode is localized on the brane
(see Fig. \ref{fig_FermionIV_Zero}), while the massless right-chiral
fermion does not exist. The class of functions (\ref{Fclass})
describe a model where both type of fermions are localized on the
brane and only the left-chiral KK spectrum possesses a massless
mode.

\begin{figure*}[htb]
\begin{center}
\includegraphics[width=7cm]{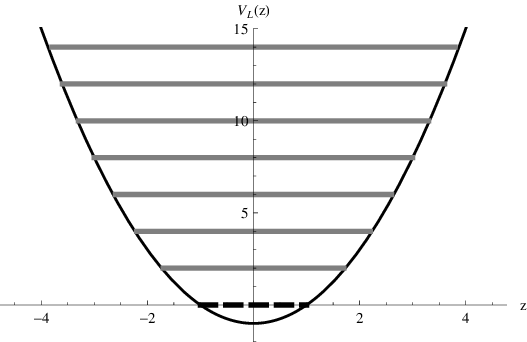}
\end{center}\vskip -5mm
\caption{ The shape of the potential $V_{L}$ is represented for
$k=0$; the dashed line corresponds to $m^2_{L_0}=0$ and the first
seven massive levels ($1\leq n\leq 7$) of the $m^2_{L_n}$ spectrum
are given by the grey lines. The parameters are set to $H=1$ and
$M=1$ in the figure.} \label{Left_potential_case4}
\end{figure*}

\begin{figure*}[htb]
\begin{center}
\includegraphics[width=7cm]{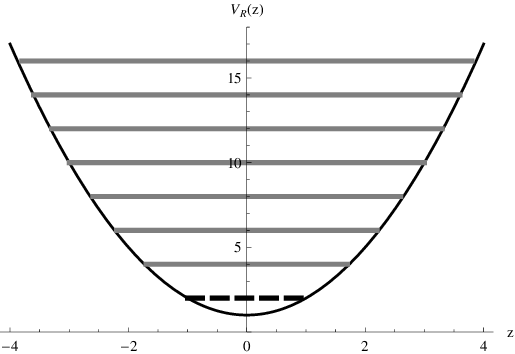}
\end{center}\vskip -5mm
\caption{ Here the shape of the potential $V_{R}$ is represented for
$k=0$; the dashed line corresponds to the massive state
$m^2_{R_0}=2MH$ and the first seven massive levels ($1\leq n\leq 7$)
of the $m^2_{R_n}$ spectrum are given by the grey lines. The
parameters are set to $H=1$ and $M=1$ in the figure.}
\label{Right_potential_case4}
\end{figure*}

Let us consider the simplest case $k=0$ in (\ref{Fclass}), the
potentials (\ref{leftpotential}) and (\ref{rightpotential}) take the
form:
\begin{subequations}
\begin{eqnarray}
  V_L(z)&=& H^{2}M^{2}z^{2}-HM, \\
  V_R(z)&=& H^{2}M^{2}z^{2}+HM.
\end{eqnarray}
\end{subequations}
Both potentials constitute shifted one--dimensional quantum harmonic
oscillator potentials and are plotted in Figs.
\ref{Left_potential_case4} and \ref{Right_potential_case4},
respectively. The corresponding Schr\"odinger--like equations for
the potentials $V_{L,R}$ can be written as:
\begin{subequations}
\begin{eqnarray}
-\partial_{z}^{2}L_n(z)+H^{2}M^{2}z^{2}L_n(z)=\left(m_{L_n}^{2} +HM\right)L_n(z), \\
-\partial_{z}^{2}R_n(z)+H^{2}M^{2}z^{2}R_n(z)=\left(m_{R_n}^{2}-HM\right)R_n(z).
\end{eqnarray}\label{Shroequa0}
\end{subequations}
Thus, the above equations describe a quantum harmonic oscillator in
one dimension if we define the effective energy modes $E_{L,R}$ and
the effective potentials $\overline V_{L,R}$  as follows
\begin{subequations}
\begin{eqnarray}
E_{L_n}=\frac{1}{2}\left( \frac{m_{L_n}^{2}}{M} +H \right) , \\
E_{R_n}=\frac{1}{2}\left( \frac{m_{R_n}^{2}}{M} -H \right),\\
\overline V_L(z)=\overline V_R(z)= H^{2}M^{2}z^{2}.
\end{eqnarray}
\end{subequations}
We see that both potentials $\overline V_{L,R}$ are equal, then,
they have the same asymptotic behavior $\overline
V_{L,R}\rightarrow\pm\infty$ when $z\rightarrow\pm\infty$ and have
their minima at $z=0$; thus, the potentials $\overline V_{L,R}$
admit the same number of bound energy modes for the left-- and
right--chiral fermions. The corresponding spectra of energies for
the equations (\ref{Shroequa0}) are given by
\begin{subequations}
\begin{eqnarray}
E_{L_n}=H\left(n+\frac{1}{2}\right)=\frac{1}{2}\left( \frac{m_{L_n}^{2}}{M} +H\right),   \\
E_{R_n}=H\left(n+\frac{1}{2}\right)=\frac{1}{2}\left(\frac{m_{R_n}^{2}}{M}-H
\right),
\end{eqnarray}\label{Shroequa}
\end{subequations}
where $n=0,1,2,\cdots$. Then, the spectra for the left and right
masses $m_{L,R}^2$ are
\begin{eqnarray}
m_{L_n}^2=2MHn \qquad {\mbox{\rm and}} \qquad m_{R_n}^2=2MH(n+1),
\end{eqnarray}\label{massesLRIV}
respectively, as indicated in Figs. \ref{Left_potential_case4} and \ref{Right_potential_case4}.

Like in the case III, we see that the KK mass spectrum for the
right--chiral fermions does not include a massless mode, while the
left--chiral fermions contains the massless KK mode. By rewriting
equations (\ref{Shroequa0}) we see that both of them represent
quantum harmonic oscillators along the fifth dimension
\begin{subequations}
\begin{eqnarray}
-\partial_{z}^{2}L_n(z)+H^{2}M^{2}z^{2}L_n(z)=E_{L_n}L_n(z), \\
-\partial_{z}^{2}R_n(z)+H^{2}M^{2}z^{2}R_n(z)=E_{R_n}R_n(z).
\end{eqnarray}\label{Shroequa1}
\end{subequations}
The zero mode can be localized for both the left-- and right--chiral
fermions and has the explicit form
\begin{subequations}
\begin{eqnarray}
L_0(z)=\frac{(HM)^{\frac{1}{2}}}{\sqrt{2\pi}}\text{e}^{-\frac{1}{2}HMz^2},\label{L04} \\
R_0(z)=\frac{(HM)^{\frac{1}{2}}}{\sqrt{2\pi}}\text{e}^{-\frac{1}{2}HMz^2},\label{R04}
\end{eqnarray}
\end{subequations}
where (\ref{L04}) is a massless bound state, while (\ref{R04}) is a
massive one.

The total spectra for the equations (\ref{Shroequa1}) is expressed
in terms of Hermite polynomials by the next eigenfunctions
\begin{subequations}
\begin{eqnarray}
L_n(z)_=\frac{B_L}{2^{n}}(HM)^{\frac{1}{4}}\text{H}_{n}(\sqrt{HM}z)\text{e}^{-
\frac{1}{2}HMz^2}, \\
R_n(z)=\frac{B_R}{2^{n}}(HM)^{\frac{1}{4}}\text{H}_{n}(\sqrt{HM}z)\text{e}^{-\frac{1}{2}HMz^2},
\end{eqnarray}
\end{subequations}
where $B_L$ and $B_R$ are normalization constants.

The spectrum for both type of fermions is discrete and the
separation between two contiguous KK modes is given by
\begin{eqnarray}
\Delta m_n=\frac{\sqrt{2HM}}{\sqrt{n+1}+\sqrt{n}}.\label{gap_m}
\end{eqnarray}
The above relations tell us  that if $n \rightarrow \infty$, $\Delta
m_n\rightarrow 0$, then, for higher KK massive modes the spectrum is
quasi-continuous.

The previous result shows that by choosing
$F(z)=Hz\text{e}^{-A(z)},$ it is possible to localize on the brane
the whole discrete KK spectrum: both massless and massive modes for
the left-chiral fermions and just the massive ones for the
right-chiral ones.

\section{Corrections to Coulomb's law}
\label{CCL}

In this Section we shall compute the Coulomb's law modification coming from the contribution of the 
massive KK modes of the vector field. This computation is similar to the one that leads to the
corrections to Newton's law coming from the contribution of the KK massive tensor modes of the metric.

In 4D Quantum Electrodynamics, the potential created by the Yukawa interaction between two fermions 
and a gauge field is given by $L_I = -e \bar{\psi}(x) \gamma^{\mu} A_{\mu}(x) \psi(x)$ with the vertex 
factor $-i e \gamma^{\mu}$ \cite{peskin}. Following this line of reasoning, within the context of our 
thick braneworld model, we shall start with fermion fields which propagate in all the five dimensions 
and we shall take into account the fact that the 4D fermion with left chirality is the zero mode of the 
5D fermion, just as we have done in the previous Section of this paper. Hence, the interaction between 
the fermions and the gauge boson reads \cite{HuangPRD2004}
\begin{eqnarray}
     S_I=  {\int} d^4 x dz~ \sqrt{-g}~ (-e_5 )\bar{\Psi}(x,z)  \Gamma^M A_M(x,z)  \Psi(x,z),
\end{eqnarray}
where $e_5$ is a 5D coupling constant. Then, after dimensional reduction of this action, all vector KK 
modes will interact with the fermion zero mode (the 4D massless fermion) which is localized on the brane:
\begin{eqnarray}
  S_I&\supset&  \sum \!\!\!\!\!\!\!\! {\int_n} {\int} d^4 x dz~\sqrt{-\hat{g}}~\text{e}^{5A}
       (-e_5)\text{e}^{-2A} \bar{\psi}_0(x) L_0(z)
       \text{e}^{-A}\gamma^{\mu} a_{\mu}^{(n)}(x)
       \text{e}^{-A/2} \rho_n(z)
       \text{e}^{-2A}\psi_0(x) L_0(z) \nonumber \\
  &=&  (-e_5 )\sum \!\!\!\!\!\!\!\! {\int_n} {\int} dz~\text{e}^{-A/2}~\rho_n(z)  L_0^2(z)
       {\int} d^4 x \sqrt{-\hat{g}}~
       \bar{\psi}_0(x)  \gamma^{\mu} a_{\mu}^{(n)}(x) \psi_0(x) \nonumber \\
  &=& {\int} d^4 x~ \sqrt{-\hat{g}} ~
     \Big\{ -e\bar{\psi}_0(x) \gamma^{\mu} a_{\mu}^{(0)}(x)  \psi_0(x)
            -\sum \!\!\!\!\!\!\!\! {\int_n}~~ \epsilon_n \bar{\psi}_0(x) \gamma^{\mu} a_{\mu}^{(n)}(x) \psi_0(x)
     \Big\},
\end{eqnarray}
where, by taking into account the equality (\ref{ZeroVector}),
$$
e=e_5 {\int} dz~\text{e}^{-A/2} ~\rho_0(z)  L_0^2(z)
   =e_5\sqrt{\frac{b}{\pi}}
        {\int} dz~ L_0^{2}(z)
   =e_5\sqrt{\frac{b}{\pi}}  
$$
is the usual 4D charge of the fermion trapped on the brane, $\epsilon_n$'s are 4D effective couplings
\begin{eqnarray}
 \epsilon_n \equiv   e_5 {\int} dz~\text{e}^{-A/2} ~\rho_n(z)  L_0^2(z)
   = e \sqrt{\frac{\pi}{b}} {\int} dz~\text{e}^{-A/2} ~\rho_n(z)  L_0^2(z),
\label{epsilon_n}
\end{eqnarray}
with $\rho_n(z)$ being the solution of the Schr\"odinger equation (\ref{SchEqVector1}) given by (\ref{gralsolnPP})
and $\sum \!\!\!\!\!\!{\int}\,_n$ stands for summation or integration (or both) with respect to $n$, depending on the 
respective discrete or continuos (or mixed) character of the $a_{\mu}^{(n)}(x)$ and 
$\epsilon_n(z)$, where the latter is determined by $\rho_n(z)$.

In the nonrelativistic limit the Coulomb potential (and its corrections) between two charged fermions is 
determined by the KK photon exchange process and turns out to be
\begin{eqnarray}
 V(r) &=& \frac{e^2}{4{\pi}r}+ \int_{m_0}^{\infty} 
 dm \frac{\epsilon_n^2}{4{\pi}r}\text{e}^{-m r}\noindent\nonumber\\
      &=& \frac{e^2}{4{\pi}r}
         \left[ 1 + \frac{\pi}{b} \int_{m_0}^{\infty} dm\ \text{e}^{-m r}
                   \left({\int} dz~\text{e}^{-A(z)/2} ~\rho_n(z)  L_0^2(z)\right)^2
         \right],
\end{eqnarray}
where $m_0=H/2$ is the first excited massive mode. Thus, the Coulomb potential and its corrections 
respectively come from the vector zero mode and the massive KK vector modes.

We can further proceed to the analytical calculation of $V(r)$ which is not an easy work, but it is still affordable. Let us compute first the 4D effective couplings $\epsilon_n$. 
We shall use the third solution for the fermionic localization mechanism with $F(z)=\sinh(Hz)$ for which the normalizable fermion zero mode is 
\begin{eqnarray}
L_{0}(z) =\left[\frac{H~\Gamma\left(\frac{b+2M}{2b}\right)}
{\sqrt{\pi}~\Gamma\left(\frac{M}{b}\right)}\right]^{\frac{1}{2}}
\text{sech}^{\frac{M}{b}}(Hz).
\end{eqnarray}
By substituting the warp factor (\ref{warpfactorz}) and the expression (\ref{gralsolnPP}) for $\rho_n$ in 
(\ref{epsilon_n}) we obtain
\begin{eqnarray}
\epsilon_n &=& e\sqrt{H}\,\frac{\Gamma\left(\frac{2M+b}{2b}\right)}{\Gamma\left(\frac{M}{b}\right)} 
{\int} dz~\text{sech}^{\frac{4M-b}{2b}}(Hz)~\left[\sum_{\pm} C_{\pm}(\beta) 
 P_{1/2}^{\pm i\beta}\left(\text{tanh}(Hz)\right)\right]\nonumber \\
   &=& e\sqrt{\frac{\pi}{H}}\,\frac{\Gamma\left(\frac{2M+b}{2b}\right)}{\Gamma\left(\frac{M}{b}\right)}
\frac{\Gamma\left(\frac{4M-b}{4b}\right)}{\Gamma\left(\frac{4M+b}{4b}\right)} 
~\left[\sum_{\pm} C_{\pm }(\beta) P_{1/2}^{\pm i\beta}\left(0\right)\right],
\label{epsilon-delta}
\end{eqnarray}
where we have used the following definition of the delta function\footnote{It is straightforward to check 
that this definition possesses all the properties of the normalized to unity delta distribution function.} 
corresponding to the thin brane limit when $H\to\infty$:
\begin{eqnarray}
\delta (z) = \lim_{H\to\infty}
~\frac{H\Gamma\left(\frac{4M+b}{4b}\right)}{\sqrt{\pi}\Gamma\left(\frac{4M-b}{4b}\right)}
\text{sech}^{\frac{4M-b}{2b}}(Hz), \qquad\qquad 4M>b,
\label{delta}
\end{eqnarray}
and $\beta=\sqrt{\frac{m^2}{H^2}-\frac{1}{4}}$.
Once we have these 4D effective couplings at hand we can write the Coulomb potential as follows
\begin{eqnarray}
 V(r) &=& \frac{e^2}{4{\pi}r}\left[1 + \frac{\pi}{H} 
\left(\frac{\Gamma\left(\frac{2M+b}{2b}\right)}{\Gamma\left(\frac{M}{b}\right)}
\frac{\Gamma\left(\frac{4M-b}{4b}\right)}{\Gamma\left(\frac{4M+b}{4b}\right)}\right)^2
\int_{m_0}^{\infty} dm\ \text{e}^{-m r} \left|\sum_{\pm} C_{\pm}(\beta) P_{1/2}^{\pm i\beta}\left(0\right)\right|^2\right]
\noindent\nonumber\\
      &=& \frac{e^2}{4{\pi}r}\left[1 + \frac{2\pi}{H} 
\left(\frac{\Gamma\left(\frac{2M+b}{2b}\right)}{\Gamma\left(\frac{M}{b}\right)}
\frac{\Gamma\left(\frac{4M-b}{4b}\right)}{\Gamma\left(\frac{4M+b}{4b}\right)}\right)^2
\int_{m_0}^{\infty} dm\ \text{e}^{-m r} 
\left|\frac{\Gamma\left(1+i\beta\right)}{\Gamma\left(\frac{1}{4}-
\frac{i\beta}{2}\right)\Gamma\left(\frac{5}{4}-\frac{i\beta}{2}\right)}\right|^2\right]. 
\noindent\label{CCl_Gammas}
\end{eqnarray}
Here we have taken into account the fact that the normalization constants for the associated Legendre
functions are given by $\left|C_{\pm}(\beta)\right|=\frac{\left|\Gamma(1+i\beta)\right|}{\sqrt{2\pi}}$, 
according to (\ref{C+C-}), as well as the following relation
\begin{eqnarray}
P_{\nu}^{\mu}(0)=
\frac{2^{\mu}\sqrt{\pi}}{\Gamma\left(\frac{1-\nu-\mu}{2}\right)\Gamma\left(1+\frac{\nu-\mu}{2}\right)}.
\label{Legendre_0}
\end{eqnarray}
Thus, according to (\ref{CCl_Gammas}) the corrected Coulomb potential can be written in the form
\begin{eqnarray}
 V(r) = \frac{e^2}{4{\pi}r}\left[1 + \Delta V\right],
\label{VCCl}
\end{eqnarray}
where the correction $\Delta V$ reads
\begin{eqnarray}
 \Delta V = 
2\pi\left[\frac{\Gamma\left(\frac{2M+b}{2b}\right)}{\Gamma\left(\frac{M}{b}\right)}
\frac{\Gamma\left(\frac{4M-b}{4b}\right)}{\Gamma\left(\frac{4M+b}{4b}\right)}\right]^2
\frac{1}{\left|\Gamma\left(\frac{1}{4}\right)\Gamma\left(\frac{5}{4}\right)\right|^2}
\frac{e^{-Hr/2}}{Hr}\left(1+{\cal O}\left(\frac{1}{Hr}\right)\right).
\label{CCptl}
\end{eqnarray}
When doing this computation, in (\ref{CCl_Gammas}) we have performed an expansion of the prefactor that 
multiplies the exponential function in the integrand with respect to $m_0=H/2$ (which corresponds to $\beta=0$)
since the corrections to the Coulomb potential are dominated by the sector of small massive KK vector modes.

Alternatively, and perhaps a more clear approach, consists in changing the integration variable in (\ref{CCl_Gammas})
from $m$ to $\beta$ and expanding the prefactor that multiplies the exponential function around $\beta=0$ since, as 
stated above, the corrections to the Coulomb potential are dominated by the small-$\beta$ region. Both integrations
yield the same result.

\section{Conclusions and discussion}
\label{SecConclusion}

In this paper, the localization and mass spectra of various bulk matter fields on a thick brane
generated by pure constant curvature in 4D and 5D, without the inclusion of scalar fields, has
been investigated.

The possibility of having a phenomenologically consistent small 4D cosmological constant in the
model
was stressed. Furthermore, all of the models considered in this work possess a mass gap in their
KK spectrum, as it also happens in the KK spectrum of the metric fluctuations, which is
proportional to $H$, and hence, proportional to the square root of the 4D cosmological constant.
Thus, the height of the mass gap, which determines the energy scale at which the KK fluctuations
can be excited, somehow is defined by the 4D cosmological constant and is big/small if the Hubble
parameter is big/small. However, for the cases III and IV of the fermion localization model, we
observe that the corresponding masses of the bound states are also multiplied by the factors $M/b$
and $M$, respectively, and directly depend on the value of these 5D parameters.

By considering the limit in which the Hubble parameter tends to zero, $H\rightarrow 0$, the energy
scale of the mass gap of the spectra of different field fluctuations also tends to vanish. Thus,
for the metric fluctuations, in this limit we recover an exponential solution for the warp factor
with a 4D metric with Poincar\'e symmetry instead of the de Sitter one, possessing a gapless
spectrum of graviton KK excitations as it happens in the Randall--Sundrum model. Mathematically,
this procedure involves the requirement of an analytic continuation of the thickness parameter
$ib\rightarrow k$ and the imposition of $Z_2$--symmetry on the relevant 5D manifold. Thus, for
consistency, a delta function source should be added to the setup, introducing as well a brane
tension in this limiting model. This brane tension can be fine-tuned with the bulk cosmological
constant in order to guarantee a vanishing induced cosmological constant on the brane \cite{rs}.
Thus, the Randall--Sundrum model was obtained as a limit of this braneworld configuration when the
Hubble parameter disappears by performing an analytic continuation of the brane thickness
parameter $b$. Strictly speaking, this limit does not form part of the original solution, however, the 
resulting solution is physically meaningful when analytically continuing the brane thickness parameter.
It would be very interesting to study in more detail the possible ``transitions" between the de Sitter and 
anti--de Sitter spacetimes in the braneworld paradigm.

We also included a mechanism which allows one to obtain TeV physical mass scales from fundamental
Planck mass scales in our model by generalizing our pure geometric thick de Sitter braneworld to
the case when the Standard Model fields are confined to a {\it positive} thin brane located at a
distance $y_2$ from the Planck brane. For this mechanism to work we need a compactification scale
of the {\it same order} of the inverse thickness parameter.

As an aside result we observe that while the gauge hierarchy is of order $10^{-15}$, the
corresponding hierarchy between the currently observed 4D Hubble parameter and the bulk inverse
thickness parameter of the model is ten orders less:
\begin{eqnarray}
\frac{M_*}{M_{Pl}}\approx 10^{-15} \qquad \mbox{vs} \qquad \frac{H}{b}\approx 10^{-5},
\label{cchierarchy}
\end{eqnarray}
a result which is closely related to the recovery of both the correct 4D gravitational couplings
and the actually observed accelerated expansion of the universe in our de Sitter braneworld.

With respect to the matter localization in the purely geometric de Sitter thick braneworld, it
turns out that for scalar and vector fields, both the scalar and vector zero modes can be
localized on the thick brane, and there exists a mass gap in the respective spectra. For massive
scalar fields, the spectrum contains at most two bound states and consists of a massless mode (the
ground state), a bound excited KK mode and a series of delocalized continuous massive KK modes.
However, if the thickness parameter is appropriately related to the mass of bulk scalar field
($m_s>\frac{\sqrt{7}}{2}b$), then there exists only one bound state corresponding to the massless
zero mode. For vector fields, the spectrum consists of a massless mode (the ground state) and a
series of continuous massive KK modes separated by a mass gap. Since the zero mode massless gauge
field is localized in our model, the Coulomb law is recovered on the brane through the photon 
exchange of two interacting 5D fermions. The corresponding corrections to Coulomb's law have been
computed as well, showing that they exponentially decay. However, in the expression for this 
correction there is a prefactor which depends on the 5D parameters of the massless zero mode
fermion field $M$ and $b$.

For the localization of a fermion zero mode, we must introduce the mass term $M F(z)\Psi\bar\Psi$
in the 5D action. Four cases have been investigated: For $F(z)= \varepsilon(z)$ and
$F(z)= \tanh(Hz)$, the fermion zero mode cannot be localized on the brane. For $F(z)=\sinh(Hz)$,
only the left--chiral fermion zero mode can be localized on the brane and there exist a mass gap
for both left-- and right--chiral fermions. The {\it finite} number of bound massive KK modes of
left-- and right--chiral fermions is the same, and it is determined by the ratio $M/b$. Hence, the
massless fermion localized on the brane consists of just the left--chiral KK mode, while the
massive fermions localized on the brane consist of left-- and right--chiral KK modes and
constitute the 4D Dirac massive fermions. For the case in which
$F(z)=(Hz)^{2k+1}\text{e}^{-A(z)}$ we qualitatively show that all the mass spectra for the left--
and right--chiral KK modes are {\it infinite}, discrete and are localized on the brane. For $k>0$
it is difficult to solve the eigenvalue problem for KK modes exactly, thus, we consider the
simplest case where $k=0$ and we found that it resembles the one--dimensional quantum harmonic
oscillator problem in which the squared mass gap $\Delta m_{L,R}^2$ between two contiguous states
for left-- and right--chiral fermions is equidistant. As we can see from the expressions for the
$m_{L_n}^{2}$ and $m_{R_n}^{2},$ all the right-chiral fermion spectrum is shifted in $2MH$ with
respect to the left-chiral fermion spectrum, while, as it is shown in (\ref{gap_m}), the mass gap
tends to vanish $\Delta m_n \rightarrow 0$ when $n\rightarrow \infty$, leading to a
quasi--continuous spectrum.

\section*{Acknowledgement}

HG and YXL were supported by the Huo Ying–Dong Education Foundation of Chinese Ministry of Education
(No. 121106), the National Natural Science Foundation of China (No. 11075065), and the Fundamental 
Research Funds for the Central Universities (No. lzujbky-2013-18 and No. 5051307001). The research of AHA, DMM and RRML 
was supported by grants  CIC-UMSNH (No. 4.16), CONACYT (No. 60060-J) and PAPIIT-UNAM (No. IN103413-3  Teor\'{i}as de Kaluza-Klein, inflaci\'on y perturbaciones gravitacionales). AHA acknowledges useful and 
illuminating discussions with Roman Linares and Ulises Nucamendi, he is grateful as well to the staff of 
the ICF, UNAM for hospitality, and thanks SNI for support. DMM acknowledges a postdoctoral grant from 
DGAPA-UNAM and RRML acknowledges a PhD grant from CONACYT.

\end{document}